\documentclass[lettersize,journal]{IEEEtran}
\usepackage{cite}
\usepackage{tikz}
\usepackage{amsmath}
\usepackage{acronym}
\usepackage{macros}
\usepackage{subcaption}
\usepackage{xspace}
\usepackage{xcolor} 
\usepackage{soul}
\usepackage{listings}
\usepackage{caption}
\PassOptionsToPackage{hyphens}{url}\usepackage{hyperref}
\usepackage{breakurl}

\acrodef{abac}[ABAC]{Attribute-Based Access Control}
\acrodef{ac}[AC]{Access Control}
\acrodef{acl}[ACL]{Access Control List}
\acrodef{ahd}[AHD]{Abnormal Health Detection}
\acrodef{ai}[AI]{Artificial Intelligence}
\acrodef{aks}[AKS]{Azure Kubernetes Service}
\acrodef{alu}[ALU]{Arithmetic Logic Unit}
\acrodef{ann}[ANN]{Artificial Neural Network}
\acrodef{api}[API]{Application Programming Interface}
\acrodef{as}[AS]{Autonomous System}
\acrodef{asn}[ASN]{Autonomous System Number}
\acrodef{aws}[AWS]{Amazon Web Services}
\acrodef{bilstm}[Bi\-LSTM]{Bidirectional Long Short-Term Memory}
\acrodef{bmv2}[BMv2]{Behavioral Model version 2}
\acrodef{bnn}[BNN]{Binary Neural Network}
\acrodef{bow}[BoW]{Bag-of-Words}
\acrodef{ca}[CA]{Certificate Authority}
\acrodef{cbc}[CBC]{Cipher Block Chaining}
\acrodef{caas}[CaaS]{Container as a Service}
\acrodef{cicd}[CI/CD]{Continuous Integration/Continuous Delivery}
\acrodef{cncf}[CNCF]{Cloud Native Computing Foundation}
\acrodef{cni}[CNI]{Container Network Interface}
\acrodef{cnn}[CNN]{Convolutional Neural Network}
\acrodef{cpe}[CPE]{Customer Premise Equipment}
\acrodef{crd}[CRD]{Custom\-Resource\-Definition}
\acrodef{cri}[CRI]{Container Runtime Interface}
\acrodef{csi}[CSI]{Container Storage Interface}
\acrodef{ddos}[DDoS]{Distributed Denial of Service}
\acrodef{dl}[DL]{Deep Learning}
\acrodef{dlp}[DLP]{Data Loss/Leakage Prevention}
\acrodef{dnn}[DNN]{Deep Neural Network}
\acrodef{dns}[DNS]{Domain Name System}
\acrodef{dos}[DoS]{Denial of Service}
\acrodef{dpi}[DPI]{Deep Packet Inspection}
\acrodef{ebpf}[eBPF]{extended Berkeley Packet Filter}
\acrodef{eks}[EKS]{Amazon Elastic Kubernetes Service}
\acrodef{ewma}[EWMA]{Exponential Weighted Moving Average}
\acrodef{faas}[FaaS]{Function as a Service}
\acrodef{fbk}[FBK]{Fondazione Bruno Kessler}
\acrodef{fl}[FL]{Federated Learning}
\acrodef{fnr}[FNR]{False Negative Rate}
\acrodef{foss}[FOSS]{Free and Open-Source Software}
\acrodef{fpr}[FPR]{False Positive Rate}
\acrodef{fpu}[FPU]{Floating Point Unit}
\acrodef{fvg}[\textsc{FedAvg}]{Federated Averaging}
\acrodef{gc}[GC]{Garbage Collector}
\acrodef{gcm}[GCM]{Galois/Counter Mode}
\acrodef{gcp}[GCP]{Google Cloud Platform}
\acrodef{gdpr}[GDPR]{General Data Protection Regulation}
\acrodef{gke}[GKE]{Google Kubernetes Engine}
\acrodef{gpu}[GPU]{Graphics Processing Unit}
\acrodef{grpc}[gRPC]{gRPC Remote Procedure Call}
\acrodef{ha}[HA]{Hardware Appliance}
\acrodef{iaas}[IaaS]{Infrastructure as a Service}
\acrodef{ics}[ICS]{Industrial Control System}
\acrodef{ids}[IDS]{Intrusion Detection System}
\acrodef{iid}[i.i.d.]{independent and identically distributed}
\acrodef{ilp}[ILP]{Integer Linear Programming}
\acrodef{iot}[IoT]{Internet of Things}
\acrodef{ietf}[IETF]{Internet Engineering Task Force}
\acrodef{ips}[IPS]{Intrusion Prevention System}
\acrodef{isp}[ISP]{Internet Service Provider}
\acrodef{jsd}[JSD]{Jensen-Shannon Distance}
\acrodef{k8s}[K8s]{Kubernetes}
\acrodef{kep}[KEP]{Kubernetes Enhancement Proposals}
\acrodef{ldap}[LDAP]{Lightweight Directory Access Protocol}
\acrodef{llm}[LLM]{Large Language Model}
\acrodef{lstm}[LSTM]{Long Short-Term Memory}
\acrodef{lucid}[\textsc{Lucid}]{Lightweight, Usable CNN in DDoS Detection}
\acrodef{mau}[MAU]{Match-Action Unit}
\acrodef{mbgd}[MBGD]{Mini-Batch Gradient Descent}
\acrodef{mips}[MIPS]{Millions of Instructions Per Second}
\acrodef{ml}[ML]{Machine Learning}
\acrodef{mlp}[MLP]{Multi-Layer Perceptron}
\acrodef{mssql}[MSSQL]{Microsoft SQL}
\acrodef{mtu}[MTU]{Maximum Transmission Unit}
\acrodef{nat}[NAT]{Network Address Translation}
\acrodef{netbios}[NetBIOS]{Network Basic Input/Output System}
\acrodef{ncca}[NCCA]{National Cybersecurity Certification Authorities}
\acrodef{nf}[NF]{Network Function}
\acrodef{nfv}[NFV]{Network Function Virtualization}
\acrodef{ngfw}[NGFW]{Next-Generation Firewall}
\acrodef{ngfws}[NGFWs]{Next-Generation Firewalls}
\acrodef{nic}[NIC]{Network Interface Controller}
\acrodef{nids}[NIDS]{Network Intrusion Detection System}
\acrodef{nn}[NN]{Neural Network}
\acrodef{nsc}[NSC]{Network Service Chaining}
\acrodef{ntp}[NTP]{Network Time Protocol}
\acrodef{of}[OF]{OpenFlow}
\acrodef{ood}[o.o.d.]{out-of-distribution}
\acrodef{oom}[OOM]{Out-of-Memory}
\acrodef{os}[OS]{Operating System}
\acrodef{ourtool}[P4DDLe]{P4-empowered Ddos detection with Deep Learning}
\acrodef{p4}[P4]{Programming Protocol-independent Packet Processors}
\acrodef{pca}[PCA]{Principal Component Analysis}
\acrodef{pess}[PESS]{Progressive Embedding of Security Services}
\acrodef{pid}[PID]{Process Identifier}
\acrodef{pisa}[PISA]{Protocol Independent Switch Architecture}
\acrodef{pop}[PoP]{Point of Presence}
\acrodef{portmap}[Portmap]{Port Mapper}
\acrodef{pqc}[PQC]{Post-Quantum Cryptography}
\acrodef{ppv}[PPV]{Positive Predictive Value}
\acrodef{ps}[PS]{Port Scanner}
\acrodef{pss}[PSS]{Pod Security Standard}
\acrodef{qoe}[QoE]{Quality of Experience}
\acrodef{qos}[QoS]{Quality of Service}
\acrodef{rbac}[RBAC]{Role-Based Access Control}
\acrodef{rnn}[RNN]{Recurrent Neural Network}
\acrodef{rpc}[RPC]{Remote Procedure Call}
\acrodef{sa}[SA]{Service Account}
\acrodef{sdn}[SDN]{Software Defined Networking}
\acrodef{sfnr}[sFNR]{system False Negative Rate}
\acrodef{sig}[SIG]{Special Interest Group}
\acrodef{ske}[SKE]{Server Key Exchange}
\acrodef{sla}[SLA]{Service Level Agreement}
\acrodef{snf}[SNF]{Security Network Function}
\acrodef{snmp}[SNMP]{Simple Network Management Protocol}
\acrodef{ssdp}[SSDP]{Simple Service Discovery Protocol}
\acrodef{svm}[SVM]{Support Vector Machine}
\acrodef{tc}[TC]{Traffic Classifier}
\acrodef{tftp}[TFTP]{Trivial File Transfer Protocol}
\acrodef{tlcp}[TLCP]{Transport Layer Cryptographic Protocol}
\acrodef{tls}[TLS]{Transport Layer Security}
\acrodef{tor}[ToR]{Top of Rack}
\acrodef{tpr}[TPR]{True Positive Rate}
\acrodef{tsp}[TSP]{Telecommunication Service Provider}
\acrodef{ttl}[TTL]{Time to Live}
\acrodef{unb}[UNB]{University of New Brunswick}
\acrodef{vm}[VM]{Virtual Machine}
\acrodef{vne}[VNE]{Virtual Network Embedding}
\acrodef{vnep}[VNEP]{Virtual Network Embedding Problem}
\acrodef{vnf}[VNF]{Virtual Network Function}
\acrodef{vpn}[VPN]{Virtual Private Network}
\acrodef{vsnf}[VSNF]{Virtual Security Network Function}
\acrodef{waf}[WAF]{Web Application Firewall}
\acrodef{wan}[WAN]{Wide Area Network}
\acrodef{xdp}[XDP]{eXpress Data Path}
\acrodefplural{pss}[PSSs]{Pod Security Standards}

\newif\ifFIXESON
\FIXESONtrue % coment this line to remove fixme comments
\ifFIXESON
\newcommand{\fixn}[2]{\fixfootnote{\textbf{#1:} #2}}
\usepackage[draft,inline,nomargin,index]{fixme}
\else 
\newcommand{\fixn}[2]{}
\usepackage[final]{fixme}
\fi
\fxsetup{theme=color,mode=multiuser}
\FXRegisterAuthor{sm}{envsm}{\color{teal}SM}    %Salvatore
\FXRegisterAuthor{rg}{envrg}{\color{purple}RG}  %Riccardo
\FXRegisterAuthor{mf}{envmf}{\color{green!80!black}MF}
\FXRegisterAuthor{hb}{envhb}{\color{blue}HBB}
\FXRegisterAuthor{lk}{envlk}{\color{red}LK}

\fxsetup{layout={footnote}}
\fxusetheme{colorsig}

\begin{document}

\title{Thou Shall Not Pass: Gatekeeping Outbound TLS Connections}

\author{Henrique B. Brum, Matteo Franzil, Riccardo Germenia, Salvatore Manfredi, Domenico Siracusa and Luis A. Dias Knob

\thanks{This work was supported by Ministero delle Imprese e del Made in Italy (IPCEI Cloud DM 27 giugno 2022 ± IPCEI-CL-0000007) and European Union (Next Generation EU).}% <-this % stops a space

\thanks{Henrique B. Brum and Matteo Franzil are with the Department of Information Engineering and Computer Science, University of Trento, 38123 Povo, Italy, and also with the Cybersecurity Center, Fondazione Bruno Kessler, 38123 Povo, Italy (e-mail: henrique.beckerbrum@unitn.it; matteo.franzil@unitn.it).}%
\thanks{Riccardo Germenia, Salvatore Manfredi and Luis A. D. Knob are with the Cybersecurity Center, Fondazione Bruno Kessler, 38123 Povo, Italy (e-mail: rgermenia@fbk.eu; smanfredi@fbk.eu; l.diasknob@fbk.eu).}%
\thanks{Domenico Siracusa is with the Department of Information Engineering and Computer Science, University of Trento, 38123 Povo, Italy (e-mail: domenico.siracusa@unitn.it).}}

\markboth{IEEE TRANSACTIONS ON INFORMATION FORENSICS AND SECURITY,~Vol.~X, 2026}%
{Brum \MakeLowercase{\textit{et al.}}: Thou Shall Not Pass: Gatekeeping Outbound TLS Connections}


\maketitle

\begin{abstract}
Despite the widespread use of \ac{tls}, its security guarantees are frequently compromised by outdated versions and misconfigurations. To analyze this problem, we collected more than 50 million \ac{tls} handshakes over a two-week period at our research institution, Fondazione Bruno
Kessler, and analyzed three server-selected parameters against the recommendations of four \ac{tls} guidelines. Our analysis shows that while the use of insecure or outdated options is minimal, it remains persistent. More importantly, servers are adopting the latest \ac{tls} advancements much faster than official guidelines can be updated to provide directives for them. These findings, combined with the difficulty of configuring \ac{tls} clients due to their ephemeral, ubiquitous and server-dependent nature, leave users vulnerable to non-standard or outright insecure connections. To address this, we present~\toolname, a real-time, network-based tool that transparently monitors handshakes, analyzes server parameters, and, based on organizational policy, reports non-compliant connections without requiring client-side modifications. Unlike Next-Generation Firewalls,~\toolname~preserves end-to-end privacy by validating only handshakes, and offers greater flexibility in defining undesired configurations. Our evaluation shows that \toolname~sustains traffic rates of up to 100 Gbps while preventing insecure connections, with an average added processing delay of 671 ns (\ac{tls} 1.3) and 795 ns (\ac{tls} 1.2) per handshake packet, making enforcement feasible at scale.
\end{abstract}


% First, guidelines often issue contradictory rules for identical values, which creates significant confusion for administrators trying to distinguish between secure and insecure configurations.

\section{Introduction}\label{sec:intro}

\IEEEPARstart{T}{ransport} Layer Security (TLS) is a suite of cryptographic protocols that provides confidentiality and integrity in network communications. Introduced in 1999 as a standardized redesign of SSL 3.0~\cite{tls10,consensus}, it has become the de facto standard for securing data over the Internet. Over time, vulnerabilities in both protocol design~\cite{beast,breach,sweet32,3shake} and implementation~\cite{heartbleed,ticketbleed,freak} prompted several updates, resulting in the release of three new versions, \ac{tls}~1.1~\cite{rfc11}, 1.2~\cite{rfc12}, and 1.3~\cite{rfc13}. The latest version, \ac{tls}~1.3, reduces the number of configurable parameters compared to its predecessors, thus narrowing the attack surface. It also removes outdated features, while improving the overall performance. Despite these enhancements, its adoption still lags behind \ac{tls}~1.2, with only 75.3\% of surveyed websites~\cite{sslpulse} supporting it. This is mainly due to legacy support for a wide set of operating systems that have already reached end-of-life but are still used.

%This burden is further increased by the discovery of additional vulnerabilities, frequently associated with legacy cryptographic algorithms, that require the disabling of specific cipher suites or the unclear requirement to enable new extensions (e.g., 3SHAKE~\cite{3shake}). 

Although the level of customization in \ac{tls}~1.2 and earlier versions was potentially beneficial, system administrators are now implicitly required to comprehend which customizations enable which features and whether they may open the system to attacks. While each deployment relies solely on the administrator's own decisions, many cybersecurity agencies have tried to set a common security threshold for websites within their jurisdiction, resulting in the issuance of technical guidelines. However, a recently published study shows that only two out of the analyzed fifty-six public administration websites follow the guidelines of their respective countries~\cite{secrypt24}. The issue is primarily attributed to systems not being promptly updated to reflect changes in guidelines or to default configurations remaining unchanged.

To examine how current server deployments align with the latest security guidelines, we collected over 50 million TLS handshakes over two weeks from our research institution, \ac{fbk}. We focused on external servers contacted by users within our institution, and evaluated three server-selected parameters—version, cipher suites, and supported groups—against four national TLS guidelines. These parameters were chosen because they are among the most critical for the security of a \ac{tls} connection. Our analysis confirms findings from previous research~\cite{secrypt24} while offering new insights. First, while the number of insecure connections with known vulnerabilities is minimal, they still persist. Second, and most importantly, we observed the widespread usage of options that are not approved by any guidelines, yet have no known vulnerabilities. This suggests that TLS servers and clients are using the latest developments, such as \ac{pqc} algorithms, faster than guidelines are updated to include directives about them.

% Our analysis confirms findings from previous research~\cite{secrypt24} while offering new insights. First, each guideline accepts vastly different values, highlighting how even cybersecurity experts struggle to standardize TLS configurations. Second, while the number of insecure connections with known vulnerabilities is minimal, they still persist. Lastly, we observed many values used in the wild that are not approved by any guidelines yet have no known vulnerabilities. This suggests that TLS servers and clients are using the latest developments, such as \ac{pqc} algorithms, faster than guidelines are updated to include directives about them.}

In order to support system administrators secure their TLS servers, the open-source community and researchers have developed tools that detect misconfigurations and provide guidance on how to fix them~\cite{testssl, tlsfuzzer, sslyze, tlsscanner, tlsassistant}. These tools reduce the burden on administrators, who no longer need to be familiar with every potential vulnerability or cryptographic nuance, although they still require manual intervention to apply corrections. Due to their ephemeral and ubiquitous nature, configuring and identifying insecure clients is even more challenging than with servers. If clients are not properly configured, they may connect to vulnerable servers and establish insecure connections, ultimately compromising both their own security and that of the broader network. Although \ac{ngfws}, such as Palo Alto \cite{paloalto2026ngfw} and SonicWall \cite{sonicwall2023ngfw}, provide some support for blocking outdated versions and insecure \ciphers, they often require full decryption or provide limited flexibility in choosing which values to block.

% If a server’s parameters do not follow the chosen policy,~\toolname~can block the connection by replacing one of the server’s messages with a TLS \emph{Alert} reporting a handshake failure, which causes the client to end the session.

To address this challenge, we present \toolname, a network-based tool that provides a flexible and effective way to protect clients from insecure servers. \toolname~requires no client-side modifications; instead, it monitors incoming handshake messages directly within the network and verifies server-selected parameters against a chosen security policy. Since it inspects only the plaintext parts of the handshake, even when using the Encrypted Client Hello (ECH) extension, our tool fully preserves end-to-end encryption. If a server’s parameters do not follow the chosen policy,~\toolname~can block the connection and prevent an insecure connection from materializing. By using XDP~\cite{ebpf2025,hoiland2018express}, \toolname~is attached directly to the network interface, allowing it to quickly parse the necessary packets without any meaningful delay to TLS and non-TLS traffic alike. Our performance evaluation shows that~\toolname~can sustain traffic rates close to 100~Gbps and handle thousands of TLS handshakes per second with minimal impact on the overall network performance. These results demonstrate that~\toolname~can transparently operate within the network while protecting all TLS clients from potentially insecure connections.

The key contributions of this paper are the following:
\begin{itemize}
    \item \textbf{50 Million TLS Handshake Dataset:} We provide two weeks of TLS handshake traffic, totaling 50,744,009 million handshakes. While certain fields have been anonymized for privacy reasons, the dataset provides realistic, ready-to-use raw network traces, which can serve as a valuable input for researchers in the field. The dataset is publicly available\footnote{\url{https://fbk.sharepoint.com/:f:/s/TLSGatekeeper/IgDBLlznj8uRQrDYWap7rhaQAYoTxe5U8ByI4eJm9XCQSTA?e=YA3Jsy}}.
    
    \item \textbf{TLS Guideline Analysis:} We analyze this dataset against four TLS guidelines issued by different cybersecurity agencies. The results demonstrate significant differences between the guidelines and a notable delay in updating them to reflect the latest developments. The source code for all our analyses and experiments is publicly available.\footnote{The dataset link contains the source code for the experiments discussed in Sections~\ref{sec:data_collection}, \ref{sec:guidelines}, and \ref{sec:tlsgatekeeper}.}
    \item \textbf{\toolname:} We developed \toolname, a network-based tool that ensures clients connect only to servers adhering to a specific TLS security policy. By operating directly within the network, \toolname can transparently enforce these requirements across all clients. Our implementation is open source.\footnote{\url{https://github.com/daisyfbk/TLSGatekeeper}} 
\end{itemize}

The remainder of this paper is organized as follows. Section~\ref{sec:background} introduces the background concepts necessary to understand the study. Section~\ref{sec:data_collection} describes the process of collecting the TLS handshake dataset, including data cleaning and anonymization, and analyzes the most common TLS options. Section~\ref{sec:guidelines} evaluates the server-selected parameters against four national TLS guidelines. Section~\ref{sec:tlsgatekeeper} presents \toolname~and discusses its performance evaluation. Finally, Section~\ref{sec:conclusions} concludes the paper.

\section{Background}\label{sec:background}

This section explains the role of TLS guidelines, and introduces the key concepts necessary to understand how TLS exchanges information to establish a session.

\subsection{TLS Guidelines}

TLS guidelines are technical documents issued by cybersecurity agencies to establish a common security baseline within a state or federation. The security of the \ac{tls}~protocol depends on numerous configurable elements that system administrators must define. These elements range from the protocol version to the permitted cryptographic algorithms, which must be selected from a list of more than one hundred options, many of which are insecure. By specifying requirements for each of these configurable elements, guidelines assist administrators in maintaining secure and compliant server configurations. 

\subsection{TLS Handshake}

The TLS protocol negotiates session parameters through the handshake subprotocol. During this stage, the client and server exchange information to authenticate each other and establish the cryptographic algorithms to be used. The main difference between the handshakes in \ac{tls}~1.3 and earlier versions is the number of required round-trips. Whereas \ac{tls}~1.2 and older require two RTTs to complete the handshake, \ac{tls}~1.3 needs only one.

Regardless of the version, the handshake begins with the exchange of two messages: \emph{Client Hello} and \emph{Server Hello}. The client initiates the connection by sending a message containing several parameters. The most relevant—and the ones analyzed in this study—are: \emph{(i)} the proposed protocol version,\footnote{In TLS 1.2 and earlier, the protocol version field is the actual version (e.g., 0x0301, 0x0302, or 0x0303). However, in TLS 1.3, the value is fixed at 0x0303, and the actual protocol version is indicated in the \texttt{supported\_versions} extension for backward compatibility.} \emph{(ii)} the list of supported cipher suites,\footnote{Each cipher suite defines a specific combination of cryptographic algorithms to be used during communication.} and \emph{(iii)} the \texttt{supported\_groups} extension.\footnote{This extension specifies the cryptographic groups supported for key exchange.}

% The server selects one of these groups to construct its \emph{Key Exchange} message.
% In TLS 1.2 and earlier, this extension was called ``elliptic_curves''.

The server responds with its \emph{Server Hello}, which includes (among other fields) the following parameters: \emph{(i)} the selected protocol version, following the same backward compatibility rule as the \emph{Client Hello}, and \emph{(ii)} the chosen cipher suite. The server’s response regarding \emph{(iii)}, the supported groups, differs between versions. In \ac{tls}~1.2 and earlier, this information is sent in a dedicated \emph{Server Key Exchange} (SKE) message, whereas in \ac{tls}~1.3, it is integrated into the \emph{Server Hello}. Once these (and other) parameters are established, the client and server can securely encrypt and exchange data.

\section{Data Collection}\label{sec:data_collection}

In this section, we detail the data collection process used to gather TLS handshake data from our institution's network.

\subsection{Setup and Methodology}

\ac{fbk}’s internal network spans multiple locations and consists of several subnets and VLANs. The traffic is highly heterogeneous: workstations and servers run various operating systems and depend on different browsers and applications. Visitors and employees connect through separate VLANs, using both personal and institution-managed devices. Additionally, a data center hosts numerous services, some of which are publicly accessible via the institution’s own \ac{as}. Although limited to only one institution, this diversity of clients makes the network a rich and representative source of TLS traffic.

While a sizable share of these devices communicate internally, our focus was on collecting data from traffic crossing the network’s edge, specifically TLS connections initiated from the internal network to external servers. Since the monitored interface for this traffic could not have TCP Segmentation Offload (TSO) enabled, and given \textit{tcpdump}’s\footnote{\url{https://www.tcpdump.org/}} lack of stateful capabilities for tracking TLS handshakes as well as \textit{tshark}’s\footnote{\url{https://www.wireshark.org/docs/man-pages/tshark.html}} inability to capture segmented handshakes, we developed a custom script\footnote{The script is available in the following repository: \url{https://github.com/daisyfbk/Capture-TLS-Handshakes/}.} for handshake collection. It monitors traffic on TCP port~443 and begins collecting TLS handshakes upon detecting a \emph{Client Hello} message. The script then follows the corresponding flow until it reaches a TLS message that is not part of the handshake, such as a \emph{Change Cipher Spec} message, an \emph{Alert}, or \emph{Application Data}. For \ac{tls}~1.3, these messages usually follow the \emph{Server Hello} directly, whereas for \ac{tls}~1.2 and earlier, they typically appear in a new packet. If, for any reason (e.g., dropped packets), these messages are not found, the script stops monitoring that flow upon receiving a TCP \textit{FIN} or \textit{RST} packet, thereby minimizing the memory footprint. Except for rare cases\footnote{For example, a \emph{Client Hello} message that is segmented in multiple packets and cannot be parsed in a single one.} or out-of-order packets, the script reliably captures complete TLS handshakes, from the initial message to the final one, even when TCP segmentation is involved.

We collected data continuously from 2025-10-13 at 00:00~UTC until 2025-10-26 at 23:59~UTC, totaling 14 days. This two-week period, spanning Monday to Sunday and including two weekends, helps ensure that the dataset is representative of the network’s typical traffic patterns. After cleaning this dataset to remove incomplete (e.g., no \emph{Server Hello}) or broken TLS handshakes (caused by the collector), the two weeks of analyzed data contained 155,066,707~packets. Of these, 50,744,009 were TLS \emph{Server Hello} packets, which formed the basis for assessing server-selected parameters.

The dataset is provided in a compressed, hourly format and can be easily decompressed and analyzed using standard tools such as \textit{tcpdump}. The dataset can be used to reproduce the results presented in this paper, and although certain fields were anonymized for privacy reasons (see the Section \ref{sec:ethical}), it remains valuable for studying TLS traffic patterns and for other related research purposes.

\subsection{Ethical Considerations}\label{sec:ethical}

Given our need to collect possibly personal data and our intention to share the dataset with the public, we worked in close collaboration with the company’s IT and legal departments. This ensured that the whole research endeavor complied with the company’s privacy policies and applicable laws.

To begin with, data collection was strictly limited to TLS handshake packets sent to or from port 443, ensuring that no information outside the scope of this research was captured. All collected data in its non-anonymized form was stored on our institution’s secure servers, with access restricted to the IT department. The analysis scripts were executed locally on the same machine where the data resided. Finally, any traces shared with the authors were already in the same anonymized format as the public dataset, with the anonymization process described next.

The dataset was anonymized to remove fields containing sensitive or personally identifiable information (PII). IP addresses were replaced with random values, MAC addresses were zeroed out, and 802.1Q VLAN tags were removed. Additionally, both the TLS Server Name Indication (SNI) field and certificates were anonymized (set to zero). These measures ensured that the dataset could not be linked to any individual or device and prevented potential server profiling, thereby safeguarding the privacy of both employees and visitors on the network.

\subsection{Data Analysis}

\begin{figure}
    \centering
    \includegraphics[width=\linewidth]{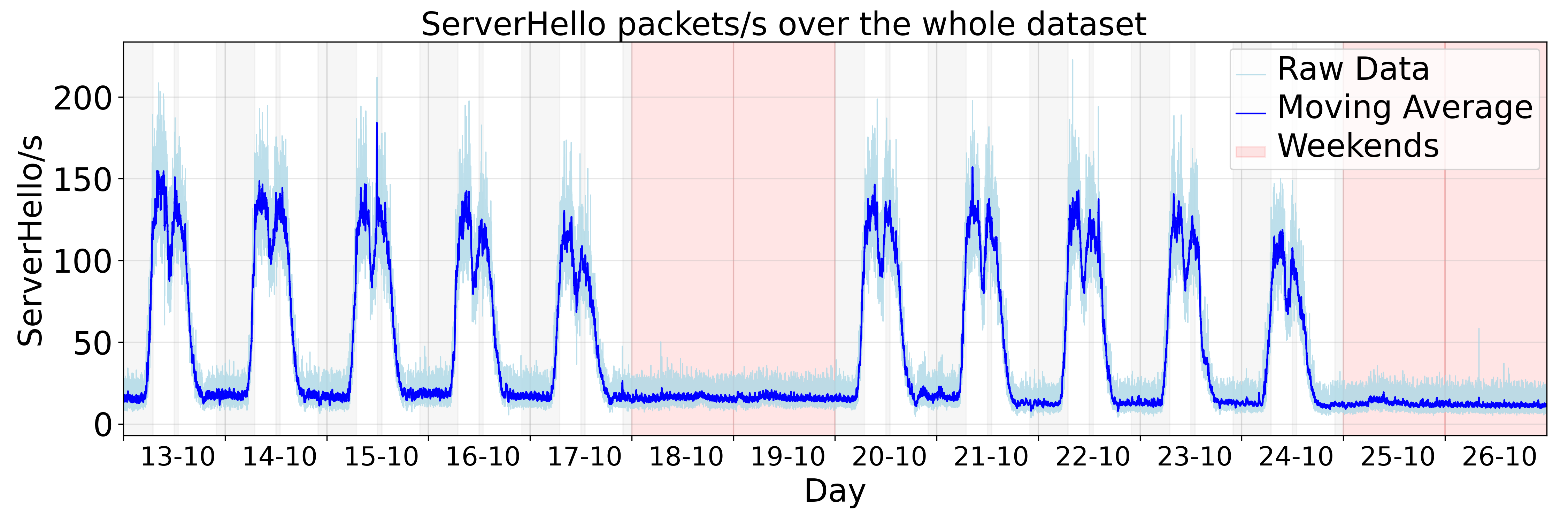}
    \caption{TLS \emph{Server Hello} packets per second over the whole data collection period.}
    \label{fig:packet_count_per_second}
\end{figure}

\begin{figure}
    \centering
    \includegraphics[width=\linewidth]{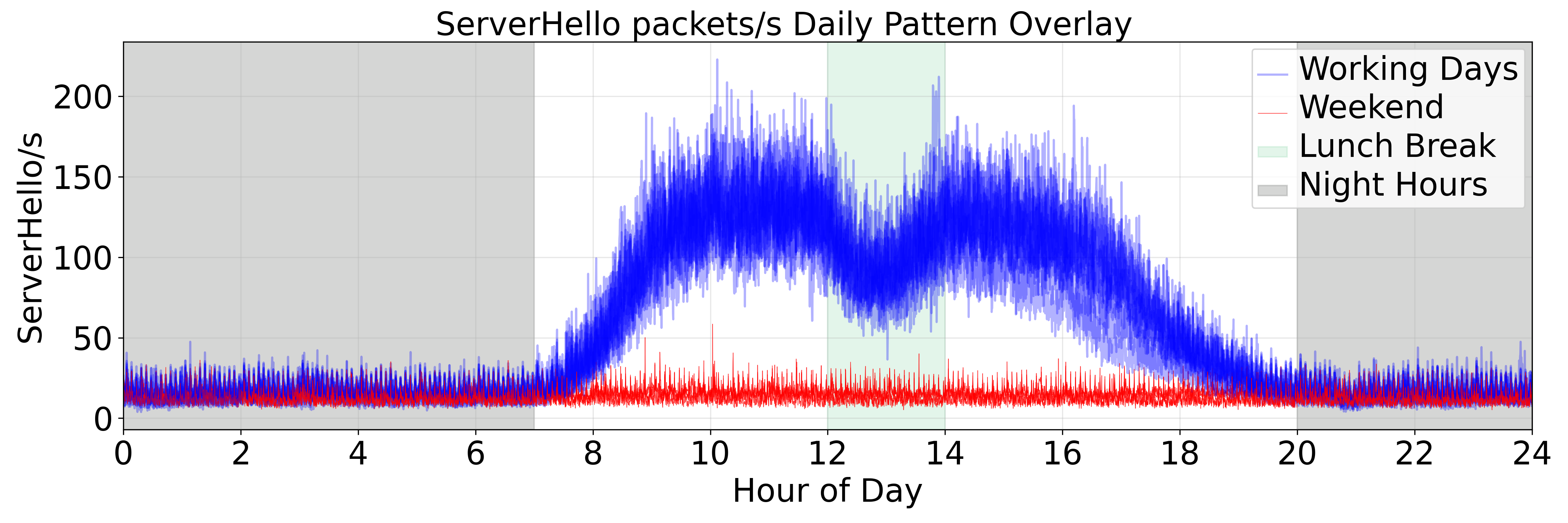}
    \caption{\emph{Server Hello} packets per second of all days overlayed.}
    \label{fig:packet_count_per_day}
\end{figure}

\vspace{5pt}\noindent\textbf{Overall trends.}\quad A deeper analysis of the collected data reveals several interesting insights. Fig.~\ref{fig:packet_count_per_second} shows the number of packets containing a \emph{Server Hello} message traversing the network per second, which corresponds to the number of TLS connections per second. A clear contrast emerges between weekdays and weekends: the former, shown on a white and gray background, exhibit substantially higher activity (during working hours) than the latter, highlighted with a red-colored background. This pattern aligns with the institution’s operational schedule, as most employees work during weekdays and within a single time zone, resulting in increased network traffic during those periods. 

%Finally, we noticed an abnormal spike around midday on 2025-10-15. For privacy reasons and since we lack non-handshake data, we could not investigate its causes, but it originated from two distinct machines contacting several hosts in a short time window.

Fig.~\ref{fig:packet_count_per_day}, which overlays all days, reveals that TLS traffic is also not evenly distributed throughout the day. There are noticeable peaks in the morning and early afternoon, a slight dip around lunchtime, and a significant drop in the evening when most employees leave the office. As the company offers a VPN service to its employees, some traffic remains during late evenings and nights, likely from remote workers. Weekend traffic is also shown in this figure, represented by the red line at the bottom. It is evident that weekend traffic is not only lower overall but also exhibits a different daily pattern. Instead of featuring peaks, the traffic remains relatively flat throughout the day, representing background activity from essential services and occasional remote access.

\begin{table}[t!]
    \centering
    \caption{TLS versions selected by the server.}
    \begin{tabular}{|l|l|}
        \hline
        TLS Version & Handshakes \\
        \hline
        TLS 1.3 & 41,130,186 \\
        TLS 1.2 & 9,546,240 \\
        FB's TLS 1.3 draft 26 & 61,899 \\
        TLS 1.0 & 5,592 \\
        TLCP: GB/T 38636-2020 & 72 \\
        TLS 1.1 & 20 \\
        \hline
    \end{tabular}
    \label{tab:tls_version_distribution}
\end{table}

\vspace{5pt}\noindent\textbf{TLS version distribution.}\quad Table~\ref{tab:tls_version_distribution} presents the distribution of \ac{tls} \textit{versions} chosen by the server in the collected dataset. The majority of packets use \ac{tls}~1.3, representing about 81\% of the total, followed by \ac{tls}~1.2, which accounts for nearly all of the remainder. Only a very small fraction of packets rely on older versions such as \ac{tls}~1.0 and \ac{tls}~1.1, both of which are deprecated and considered insecure. In addition, a small subset of packets did not match any standardized version but were still valid handshakes. These included \ac{tlcp}, a Chinese TLS variant, and Facebook’s draft version of \ac{tls}~1.3, which we further analyze in Section~\ref{sec:guidelines:anomalous_values}.

\begin{table}
    \centering
    \caption{Top 10~\ciphers~selected by servers.}
    \footnotesize
    \begin{tabular}{|l|l|}
        \hline
        Cipher Suite & Handshakes \\
        \hline
        TLS\_AES\_256\_GCM\_SHA384 & 24,132,032 \\
        TLS\_AES\_128\_GCM\_SHA256 & 16,636,426 \\
        TLS\_ECDHE\_RSA\_AES\_256\_GCM\_SHA384 & 5,632,884 \\
        TLS\_ECDHE\_RSA\_AES\_128\_GCM\_SHA256 & 2,088,654 \\
        TLS\_ECDHE\_ECDSA\_AES\_128\_GCM\_SHA256 & 766,922 \\
        TLS\_ECDHE\_ECDSA\_AES\_256\_GCM\_SHA384 & 697,728 \\
        TLS\_CHACHA20\_POLY1305\_SHA256 & 423,627 \\
        TLS\_ECDHE\_RSA\_CHACHA20\_POLY1305 & 135,149 \\
        TLS\_RSA\_AES\_256\_GCM\_SHA384 & 98,226 \\
        TLS\_RSA\_AES\_256\_CBC\_SHA1 & 47,130 \\
        \hline
    \end{tabular}
    \label{tab:cipher_suite_distribution}
\end{table}

\vspace{5pt}\noindent\textbf{Cipher suites distribution.}\quad Table~\ref{tab:cipher_suite_distribution} presents the 10 most frequently selected~\textit{\ciphers} by servers. Given that the majority of the traffic corresponds to \ac{tls}~1.3, the two most common entries are cipher suites from this version. They are followed by \ac{tls}~1.2 cipher suites, completing the top five, and representing 97\% of all observed ciphers. Across these five most common~\ciphers, there is a clear preference for \texttt{AES} in \texttt{\ac{gcm}} combined with at least \texttt{SHA-256} for hashing, reflecting secure and modern cryptographic choices. With the exception of the \texttt{TLS\_RSA\_AES\_256\_CBC\_SHA1}~\cipher, which is vulnerable to various attacks due to its reliance on \texttt{SHA1} and the \texttt{\ac{cbc}} mode of operation, the remaining four~\ciphers~use secure cryptographic primitives.

\begin{table}[t!]
    \centering
    \caption{Top 10 supported groups selected by servers.}
    \begin{tabular}{|l|l|}
        \hline
        Supported Group & Handshakes \\
        \hline
        X25519 / Curve25519 & 25,858,638 \\
        secp384r1 / P-384 & 13,591,931 \\
        X25519MLKEM768 & 6,845,778 \\
        secp256r1 / P-256 & 2,843,606 \\
        No group & 1,278,804 \\
        secp521r1 / P-521 & 279,376 \\
        X25519Kyber768\_Draft00 & 45,658 \\
        curveSM2 & 110 \\
        X25519Kyber768\_Exp & 44 \\
        X448 / Curve448 & 32 \\
        %ffdhe3072 & 32 \\
        \hline
    \end{tabular}
    \label{tab:supported_groups_distribution}
\end{table}

\vspace{5pt}\noindent\textbf{Supported groups distribution.}\quad Lastly, Table~\ref{tab:supported_groups_distribution} shows the 10 most frequently selected \emph{supported groups} by servers. At the top of the list is \texttt{Curve25519}, which is widely adopted and supported by all major browsers. The second, fourth, and sixth entries correspond to well-known prime curves standardized by NIST, offering a strong level of security. The third, seventh, and ninth entries correspond to \ac{pqc} groups, while the tenth, \texttt{curveSM2}, refers to a Chinese national group. Both categories are analyzed in detail in Section~\ref{sec:guidelines:anomalous_values}.

The fifth entry, \texttt{No group}, refers to server messages from \ac{tls}~1.2 and earlier versions, as well as \ac{tlcp}, that do not include a supported group parameter. This occurs in three scenarios: \emph{(i)} handshakes that reuse existing keys and therefore omit the \emph{SKE} message; \emph{(ii)} \emph{Server Hello} messages that omit \emph{SKE} entirely because the server certificate provides sufficient information; and \emph{(iii)} \emph{SKE} messages that specify the parameters for a cryptographic group directly, rather than relying on a predefined \texttt{named\_curve} (a behavior deprecated by RFC8422~\cite{rfc8422}, published in 2018). Finally, at the tenth position, \texttt{Curve448} was introduced in the same RFC~\cite{rfc7748} as \texttt{Curve25519}. Although \texttt{Curve448} offers strong security guarantees, its adoption has been limited, as \texttt{Curve25519} provides comparable security with superior performance.

\subsection{Discussion}
The collected dataset and its subsequent analysis provide valuable insights into current TLS server configurations and the security implications of these choices. First, \ac{tls}~1.3 has seen widespread adoption since its release nearly eight years ago, yet \ac{tls}~1.2 connections remain common. Moreover, the presence (even if minimal) of deprecated or non-standard versions is particularly problematic, as these versions contain known vulnerabilities or have not been extensively tested by the security community. Second, although the use of outdated and insecure \ciphers~is limited, their continued presence remains concerning and calls for decisive action to eliminate their use. Finally, the observed supported groups exhibit no publicly known vulnerabilities, which is encouraging, while the inclusion of \ac{pqc} algorithms indicates that some servers are preparing for future security requirements.

\section{TLS Guideline Analysis}\label{sec:guidelines}

This section analyzes the two-week dataset through the lens of four different TLS guidelines.

\subsection{Guideline Analysis}\label{sec:guidelines:guideline_compliance}

\subsubsection{Overview}

\begin{figure}
    \centering
    \includegraphics[width=\linewidth]{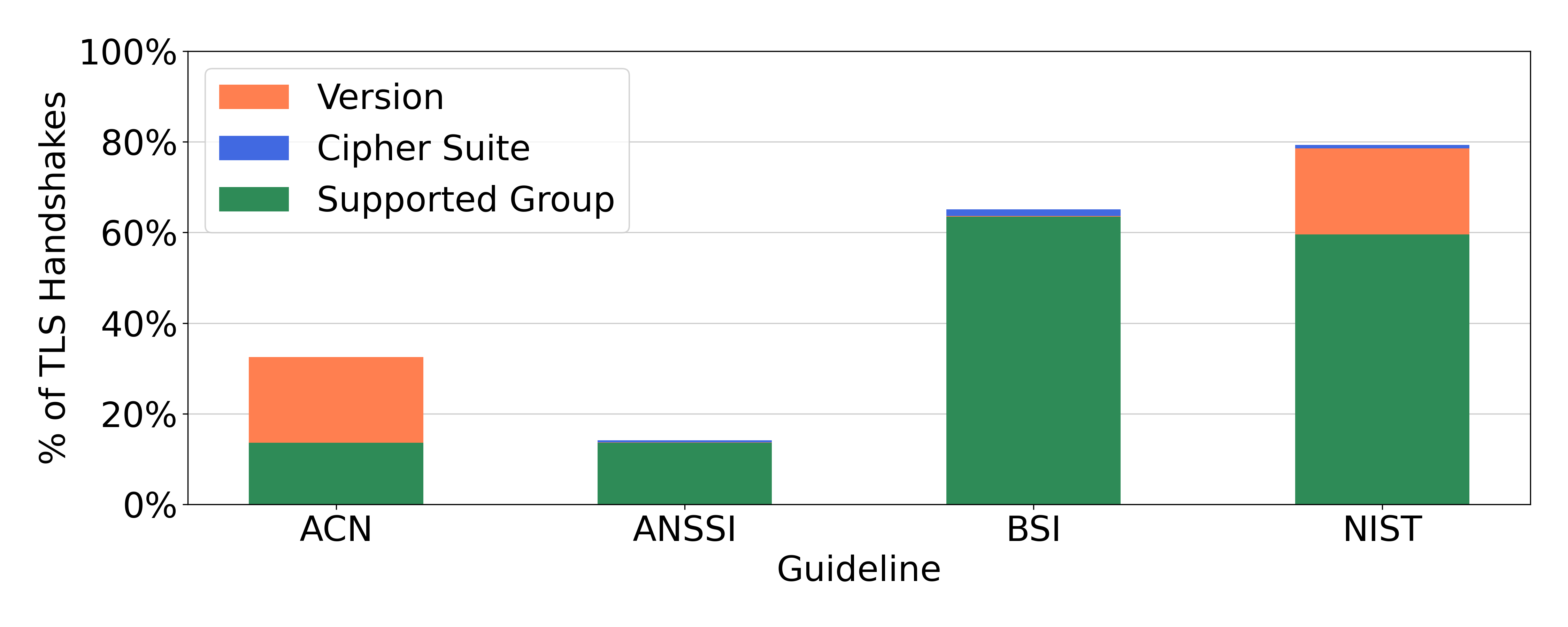}
    \caption{Amount of non-compliant TLS handshakes by guideline and parameter.}
    \label{fig:guideline_overall_compliance}
\end{figure}

We analyzed the entire dataset against four national security guidelines provided by the \datasetname~\cite{tls-compliance-dataset}: ACN (Italy), ANSSI (France), BSI (Germany), and NIST (U.S.).\footnote{To use the latest available updates, we performed our analysis on the currently unreleased "ACN" branch of the dataset, available at \url{https://github.com/stfbk/tls-compliance-dataset/tree/ACN}, commit \texttt{2677b1a}.} \footnote{We excluded the Mozilla guideline because it is not issued by a government agency, and the ENISA guideline as it focuses on security certifications rather than web server configuration.} Appendix \ref{apx:tls_guidelines} provides additional details about this dataset and our methodology. This analysis aims to compare how these guidelines differ regarding three server-selected parameters—version, cipher suite, and supported group—and establish a ground truth for comparison. For each guideline, we compared the parameters from all TLS handshakes against the values accepted by that specific guideline. Fig.~\ref{fig:guideline_overall_compliance} illustrates the percentage of non-compliant\footnote{In the context of this analysis, ``compliance" or ``compliant" refers to a server using a guideline's recommended values for the three aforementioned parameters. This does not imply that a server is fully compliant with the guideline; such a task would require a direct assessment of the server, while our analysis is based solely on handshake messages.} TLS handshakes, categorized by parameter, for all evaluated guidelines.

The first notable observation is the wide variation in compliance levels: for ANSSI, only roughly 14\% of the handshakes used non-compliant values. On the other hand, for NIST, the rate reached 80\%. Despite the role of guidelines as documents defining a clear and unified baseline security requirement, the differences among these four guidelines highlight the absence of a common consensus across countries on what is deemed secure.

The use of non-compliant supported groups is the primary cause of non-compliance across all guidelines, followed by the protocol version (the leading cause for ACN) and, at a distant third, the cipher suites. \texttt{Curve25519}, a curve widely adopted by all major browsers, is the main reason for the high number of non-compliant supported groups in the BSI and NIST guidelines. Despite its widespread usage, the guidelines do not provide an explicit rationale for excluding \texttt{Curve25519}, reflecting the general absence of a clear threat model. Another frequent source of non-compliance across all guidelines, discussed in Section~\ref{sec:guidelines:anomalous_values}, involves PQC algorithms, which have gained significant attention in recent years due to ongoing developments and the potential impact of quantum computing on current cryptographic methods.

Under guidelines that still allow \ac{tls}~1.2 (ACN and BSI), the proportion of non-compliant versions is relatively low due to the limited presence of deprecated versions such as \ac{tls}~1.0 or non-standard variants like Facebook’s \ac{tls}~1.3 draft. Although negligible compared to the total number of valid packets, their presence remains concerning, as these versions are considered insecure or, in the case of non-standard variants, lack formal approval mechanisms and may contain undisclosed vulnerabilities. Conversely, for guidelines that do not support version~1.2, the number of non-compliant connections increases, as roughly 20\% of the traffic still uses \ac{tls}~1.2.

Finally, cipher suites are not a significant source of non-compliance, with the most prevalent non-compliant cipher, \texttt{TLS\_CHACHA20\_POLY1305\_SHA256}, appearing in less than 1\% of all connections. This high level of compliance is exemplified by the fact that the five most frequently used cipher values (as shown in Table~\ref{tab:cipher_suite_distribution}) are compliant with all four guidelines and together represent 97\% of all Server Hellos. The low percentage of non-compliant cipher suites suggests that most TLS servers already employ cipher configurations that are both secure and compliant with the evaluated guidelines.

\begin{figure}[t!]
    \centering
    \includegraphics[width=\linewidth]{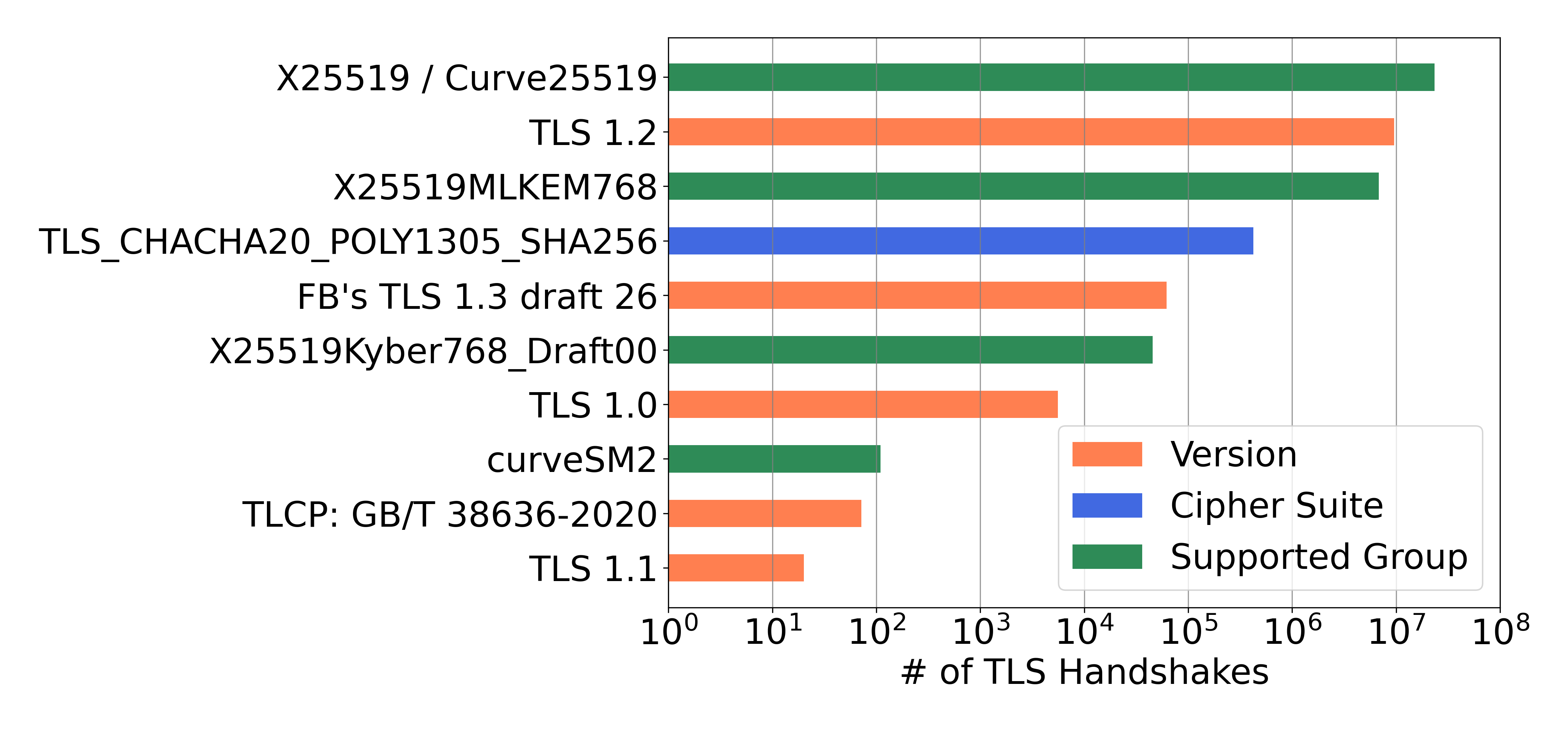}
    \caption{Top 10 non-compliant TLS values for the NIST guideline.}
    \label{fig:nist-values}
\end{figure}

\begin{table*}[t!]
    \centering
    \caption{TLS values that caused a handshake to be non-compliant but were never mentioned in any of the guidelines.}
    \footnotesize
    \begin{tabular}{|l|l|l|l|}
        \hline
        Value                                    & Parameter  & Category                    & Handshakes   \\
        \hline
        X25519MLKEM768                               & Supported Group & \ac{pqc}                            & 6,845,778 \\
        FB's TLS 1.3 draft 26                        & Version         & Facebook's TLS library              & 61,899    \\
        X25519Kyber768\_Draft00                      & Supported Group & \ac{pqc}                            & 45,658    \\
        TLS\_RSA\_WITH\_RC4\_128\_SHA                & Cipher Suite    & Insecure ciphers                    & 617       \\
        TLS\_DH\_anon\_WITH\_RC4\_128\_MD5           & Cipher Suite    & Insecure ciphers                    & 234       \\
        curveSM2                                     & Supported Group & Chinese TLS alternative             & 110       \\
        TLS\_RSA\_WITH\_CAMELLIA\_128\_CBC\_SHA      & Cipher Suite    & Insecure ciphers                    & 86        \\
        TLS\_RSA\_WITH\_CAMELLIA\_128\_CBC\_SHA256   & Cipher Suite    & Insecure ciphers                    & 85        \\
        TLS\_RSA\_WITH\_ARIA\_128\_GCM\_SHA256       & Cipher Suite    & Insecure ciphers                    & 85        \\
        TLS\_DHE\_RSA\_WITH\_CAMELLIA\_128\_CBC\_SHA & Cipher Suite    & Insecure ciphers                    & 85        \\
        TLS\_DHE\_RSA\_WITH\_CAMELLIA\_256\_CBC\_SHA & Cipher Suite    & Insecure ciphers                    & 84        \\
        TLS\_RSA\_WITH\_CAMELLIA\_256\_CBC\_SHA256   & Cipher Suite    & Insecure ciphers                    & 84        \\
        TLS\_RSA\_WITH\_CAMELLIA\_256\_CBC\_SHA      & Cipher Suite    & Insecure ciphers                    & 83        \\
        TLS\_RSA\_WITH\_ARIA\_256\_GCM\_SHA384       & Cipher Suite    & Insecure ciphers                    & 83        \\
        TLCP: GB/T 38636-2020                        & Version         & Chinese TLS alternative             & 72        \\
        ECC\_SM4\_CBC\_SM3                           & Cipher Suite    & Chinese TLS alternative             & 72        \\
        X25519Kyber768\_Exp                          & Supported Group & Facebook's TLS library & 44        \\
        TLS\_ECDH\_anon\_WITH\_NULL\_SHA                  & Cipher Suite    & Insecure ciphers                    & 11        \\
        \hline
    \end{tabular}
    \label{tab:anomalous_values}
\end{table*}

\subsubsection{NIST}\label{sec:guidelines:nist}

From these four guidelines, we selected NIST to better understand the main non-compliant values, as it exhibited the highest number of non-compliant handshakes. The NIST guideline was created as a security baseline for U.S. federal departments and agencies, but it is also useful to companies and other organizations. Figure~\ref{fig:nist-values} presents the top 10 non-compliant values according to the NIST guideline. Among these, we highlight:

\begin{enumerate}
    \item The supported group \texttt{Curve25519}, which, although not approved by NIST, is widely used and accepted by all major browsers.

    \item Although TLS 1.2 has not yet been formally deprecated, several recommendations advocate its replacement with TLS 1.3, which provides stronger security guarantees and support for PQC algorithms.

    \item The third, fifth, sixth, eighth, and ninth values are not mentioned by NIST or by any of the other evaluated guidelines. This is not surprising, as guidelines typically present only a subset of values that they consider secure and recommend for use. These values are discussed separately in Section~\ref{sec:guidelines:anomalous_values}.

    \item \texttt{TLS\_CHACHA20\_POLY1305\_SHA256} is currently discouraged by NIST, despite being cryptographically state-of-the-art.
\end{enumerate}

\subsection{Anomalous Values}\label{sec:guidelines:anomalous_values}

TLS guidelines define what values must be used and which ones are discouraged. However, they do not necessarily cover all the possible values that can be used in a TLS connection. Indeed, new ones are often standardized after the guidelines are published, or are outright tested in the wild. During our analysis, we observed that some configurations that caused a connection to be classified as non-compliant were not explicitly addressed in any of the guidelines.\footnote{This list is not exhaustive. Since compliance was evaluated sequentially (version, then \cipher, and lastly the supported group), the evaluation stopped at the first non-compliant parameter, which was the only one analyzed.} These configurations are listed in Table~\ref{tab:anomalous_values}.

We divided the anomalous values in Table~\ref{tab:anomalous_values} into four categories: \textbf{insecure~\ciphers}, \textbf{\ac{pqc}}, \textbf{Facebook's TLS library}, and a \textbf{Chinese TLS alternative}.

\vspace{5pt}\noindent\textbf{Insecure~\ciphers.}\quad Out of the 11~\ciphers~ in the ``Insecure ciphers`` category, six use the \texttt{SHA1} algorithm. This algorithm has been considered insecure since 2005 when a theoretical collision attack was discovered~\cite{sha1attack}. In recent years, attacks on \texttt{SHA1} have evolved from theoretical to practical~\cite{leurent2019collisions, leurent2020sha}. Accordingly, it was deprecated by NIST in 2011, and the agency is also planning to have it completely removed from all applications by the end of 2030~\cite{nistsha1}. Even though they do not use \texttt{SHA1}, the ciphers \texttt{TLS\_RSA\_WITH\_CAMELLIA\_128\_CBC\_SHA256} and \texttt{TLS\_RSA\_WITH\_CAMELLIA\_256\_CBC\_SHA256} rely on \texttt{\ac{cbc}}, which may be vulnerable to the LUCKY13 attack~\cite{lucky13}. However, the risk associated with these ciphers is mitigated when the “encrypt-then-MAC” \ac{tls} extension is used, as it helps defend against this vulnerability.

Additionally, two other ciphers use the \texttt{ARIA} encryption algorithm in \texttt{\ac{gcm}}, for which no vulnerabilities are currently known. The main issue with these ciphers lies in their use of the \texttt{RSA} algorithm for key exchange, which lacks the forward secrecy property. Forward secrecy ensures that even if a private key is compromised, past and future sessions remain secure. While the absence of forward secrecy is not an attack vector in itself, several guidelines recommend preferring algorithms that provide this property due to the enhanced security they offer.

Lastly, the presence of anonymous ciphers, \texttt{TLS\_ECDH\_anon\_WITH\_NULL\_SHA} and \texttt{TLS\_DH\_anon\_WITH\_RC4\_128\_MD5}, is particularly concerning. The use of these ciphers is strongly discouraged, as they omit the identity validation step, leaving connections vulnerable to Man-in-the-Middle (MITM) attacks.

%li conosciamo ma perché sono ancora in giro? (nel senso di: quale client obsoleto li può ancora usare? Da quanto sono deprecati e che problemi possono causare)
% Update a riguardo di RG: ho chiesto a RLongo e mi ha detto che teoricamente CAMELLIA e ARIA non sono deprecati/insicuri ma semplicemente poco usati all'infuori dei paesi che li hanno fatti. Come motivo per cui non appaiono si può comunuqe usare il fatto che non presentino un kex effimero e quindi non forniscono forward secrecy

\vspace{5pt}\noindent\textbf{PQC.}\quad Due to rapid advancements in quantum computing research, security researchers are increasingly concerned about the potential for quantum computers to break the public-key cryptographic algorithms currently in use. Although no existing quantum computer is capable of threatening encrypted communications, it is important to begin transitioning to post-quantum alternatives as soon as possible. This urgency arises from the risk of adversaries collecting encrypted traffic today with the intent of decrypting it in the future, a tactic known as \textit{harvest now, decrypt later}. In 2016, NIST issued a call for proposals with the goal of identifying quantum-secure alternatives to the various primitives employed in public-key cryptography. By 2024, three of the proposed algorithms, namely \texttt{ML-KEM}, \texttt{ML-DSA}, and \texttt{SLH-DSA}, were standardized by NIST~\cite{nistpqc}.

The \textit{harvest now, decrypt later} threat is a serious concern for many companies, including Cloudflare, which has been actively researching \ac{pqc} since 2017. Cloudflare recently added support for hybrid key-exchange algorithms such as \texttt{X25519MLKEM768} and \texttt{X25519Kyber768\_Draft00} (the latter now considered obsolete) in its products~\cite{cloudfarepqc}. The implementations of different post-quantum algorithms are provided by the Open Quantum Safe (OQS) project~\cite{oqs}, which integrates them into protocols and applications. Thanks to the OQS project, OpenSSL has supported the three algorithms standardized by NIST since version 3.5~\cite{opensslpqc}. In addition to these efforts, both Mozilla and Google updated their browsers (Firefox and Chrome, respectively) in 2024 to support \texttt{ML-KEM} algorithms~\cite{cloudfarepqc}. Table~\ref{tab:anomalous_values} reflects these developments, showing \texttt{X25519MLKEM768} and \texttt{X25519Kyber768\_Draft00} among the anomalous values that appeared the most in our dataset.

Although the first round of NIST standardization resulted in three selected algorithms, there is currently no definitive NIST standard regulating the use of \ac{pqc} algorithms. The ACN guidelines acknowledge this, stating that no formal recommendation is made at this time and that the document will be updated once a final standard becomes available~\cite{acn_guideline} (Section 5.5 in ACN'S document). This indicates that despite the lack of regulations at the time of writing, these algorithms (or similar ones) are likely to be integrated into guidelines in the near future.

%wow quindi la gente è interessata e sono già usati in-the-wild. CLoudlfare/Google e altri? grazie a open quantum esiste, ora con OpenSSL 3.5 sarà sempre più comune

\vspace{5pt}\noindent\textbf{Facebook's TLS library.}\quad Facebook was among the companies that contributed to the standardization of \ac{tls}~1.3 together with the \ac{ietf}. While contributing to this effort, Facebook developed an open-source library called Fizz, designed to provide a fast and secure \ac{tls}~implementation for its products~\cite{fizznews}. One of the most important differences between Fizz and other TLS libraries (e.g., OpenSSL) is that it does not support the standardized version of the \ac{tls}~1.3 protocol~\cite{rfc13}. Instead, it only implements three draft versions: \textit{draft-ietf-tls-tls13-28}, \textit{draft-ietf-tls-tls13-26}, and \textit{draft-ietf-tls-tls13-23}, with the first two being wire-compatible with RFC8446~\cite{fizz}. In addition to providing the ``standard'' version of these drafts, Fizz also includes their Facebook-specific variants of drafts 23 and 26. These variants can be identified by their use of the hexadecimal identifier \texttt{$0xfb00~||~0x{draft\_version}$} instead of the \texttt{$0x7f00~||~0x{draft\_version}$} format specified in the drafts.

Since 2024, Fizz has added support for hybrid key-exchange algorithms based on \texttt{ML-KEM} through integration with liboqs.\footnote{The Open Quantum Safe library.} An interesting detail in this integration is the introduction of two supported group algorithms with values \texttt{$0xfe00$} and \texttt{$0xfe01$}, which are reserved for private use in the IANA registry~\cite{tlsparameters}. In the Fizz source code, these values correspond to \texttt{x25519\_kyber768\_experimental}\footnote{Also referred as X25519Kyber768\_Exp.}~and \texttt{x25519\_kyber512\_experimental}, respectively, with comments indicating that they are used to ``get clean data for external traffic experiments.'' Our analysis shows that Facebook's \ac{tls}~1.3 variant is still in use, ranking as the third most common version in our dataset (see Table \ref{tab:tls_version_distribution}) and indicating that its usage remains notable.

%\footnote{Despite not being based on the latest version of the specification, that followed draft 28, the library's documentation states that its implementations of drafts 26 and 28 are wire-compatible with the final specification~\cite{fizz}.}
% Fizz currently supports TLS 1.3 drafts 28, 26 (both wire-compatible with the final specification), and 23. All major handshake modes are supported, including PSK resumption, early data, client authentication, and HelloRetryRequest.

% c'è della storia pubblica? Meta ne parla o ha pubblicato whitepaper? In cosa si differenzia, se lo fa?
% https://engineering.fb.com/2024/05/22/security/post-quantum-readiness-tls-pqr-meta/ 
% https://engineering.fb.com/2018/08/06/security/fizz/

\vspace{5pt}\noindent\textbf{Chinese TLS alternative.}\quad In 2020, the Chinese government released a standard document, titled GB/T 38636-2020, which describes a protocol functionally similar to \ac{tls}~1.2 named \ac{tlcp}.\footnote{\url{https://openstd.samr.gov.cn/bzgk/gb/newGbInfo?hcno=778097598DA2761E94A5FF3F77BD66DA}} The two protocols differ in two aspects. First, \ac{tlcp} uses two separate certificates, one for authentication and the other for encryption. Second, \ac{tlcp} only uses cryptographic algorithms from the \texttt{ShāngMì(SM)} family of cryptographic primitives. Three of these primitives are used: \texttt{SM2}, an elliptic curve–based digital signature algorithm; \texttt{SM3}, a cryptographic hash function; and \texttt{SM4}, a block cipher for symmetric encryption. In 2021, RFC8998 was published, describing how to use \texttt{SM} primitives with \ac{tls}~1.3~\cite{rfc8998}. An interesting aspect of the traffic obtained in Section~\ref{sec:data_collection}~is that the packets in which \texttt{curveSM2} was observed use the \texttt{TLS\_AES\_256\_GCM\_SHA384}~\cipher~instead of the \texttt{SM4} \cipher~described in RFC8998.\footnote{Starting with \ac{tls}~1.3, the~\ciphers~include only the encryption and hashing algorithms.} This combination of international and \texttt{SM} algorithms is problematic: it limits the set of compatible clients that can connect to the service while also failing to achieve compliance with Chinese standards.

\subsection{Discussion} 

%Although we evaluate the compliance of server-selected parameters with their corresponding guidelines, the primary goal of this analysis is to understand which parameters are recommended by the guidelines, the supposed security baseline, and how these recommendations relate to the parameters currently deployed in practice.

While TLS guidelines define a secure baseline, our analysis reveals significant differences between them, highlighting the challenge of standardizing configurations and defining which values must be used. Furthermore, while guidelines are not expected to justify every decision or omission, their limited guidance on emerging values, specifically \ac{pqc} supported groups, shows how they struggle to keep pace with recent developments.

Even tough we cannot completely evaluate a server's compliance to a guideline due to the limited visibility into server configurations during the handshake, this analysis underscores how recommended configurations can differ significantly from those used in practice. These results provide organizations with a clearer understanding of today’s TLS ecosystem: which configurations are widely used, which are discouraged for valid security reasons, and which are omitted by guidelines despite having no known vulnerabilities. Ultimately, this allows organizations to make more informed, context-aware decisions when configuring their servers and clients.

\begin{figure*}[t!]
    \centering
    \includegraphics[width=\textwidth]{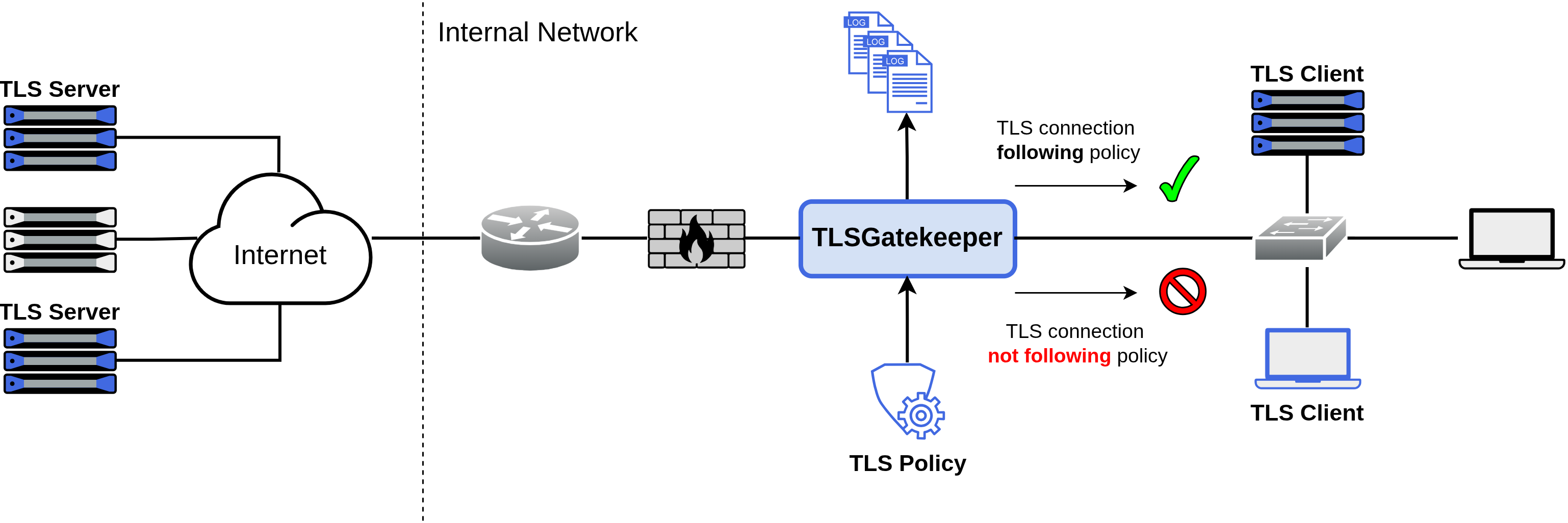}
    \caption{\toolname~deployment scenario.}
    \label{fig:deployment}
\end{figure*}

\section{TLSGatekeeper}\label{sec:tlsgatekeeper}

The prevalence of servers using outdated or insecure options, along with the need for organizations to follow specific TLS guidelines, motivated the development of \toolname. Our tool ensures that TLS connections from the internal network to external servers adhere to a defined policy for versions, \ciphers, and supported groups. This policy can be based on national guidelines or customized requirements, guaranteeing that users connect only to servers deemed secure. Fig. \ref{fig:deployment} illustrates \toolname’s operation and placement. It is deployed as a middlebox between incoming external traffic and the clients inside the institution’s network.

As a middlebox, \toolname~enforces compliance across the entire network without requiring any client-side configuration. When a client initiates a TLS connection to an external server, \toolname~inspects the server's handshake messages. Since the tool focuses on the preferences expressed in the \emph{Server Hello} message, the presence of the ECH extension does not affect validation.\footnote{The ECH extension protects privacy by hiding the SNI and other fields of the \emph{Client Hello} message.} If the server messages contain values that do not conform to the selected policy, \toolname~logs the violation and can optionally block the connection. Although blocking can be restrictive, administrators can use it to harden network security without manually filtering insecure connections. In this way, it works similarly to a firewall or a Network Intrusion Prevention System. Conversely, handshakes that align with the policy's values are authorized to pass, materializing the connection.

Existing NGFWs from Palo Alto \cite{paloalto2026ngfw}, Google Cloud \cite{google2026ngfw}, and SonicWall \cite{sonicwall2023ngfw} support blocking TLS connections that use outdated versions or weak ciphers. However, \toolname~differs in several key ways. First, \toolname~focuses only on the server-selected parameters shared in plaintext and does not require decrypting the session, therefore minimizing the processing overhead and resource consumption. Moreover, it is not restricted to a pre-determined set of values; instead, it allows system administrators complete freedom to block even the latest (and possibly non-validated) TLS \ciphers~and supported groups. Finally, it is the only tool that provides filtering capabilities for the supported groups parameter.

\subsection{Implementation}

\begin{figure}[t!]
    \centering
    \includegraphics[width=\linewidth]{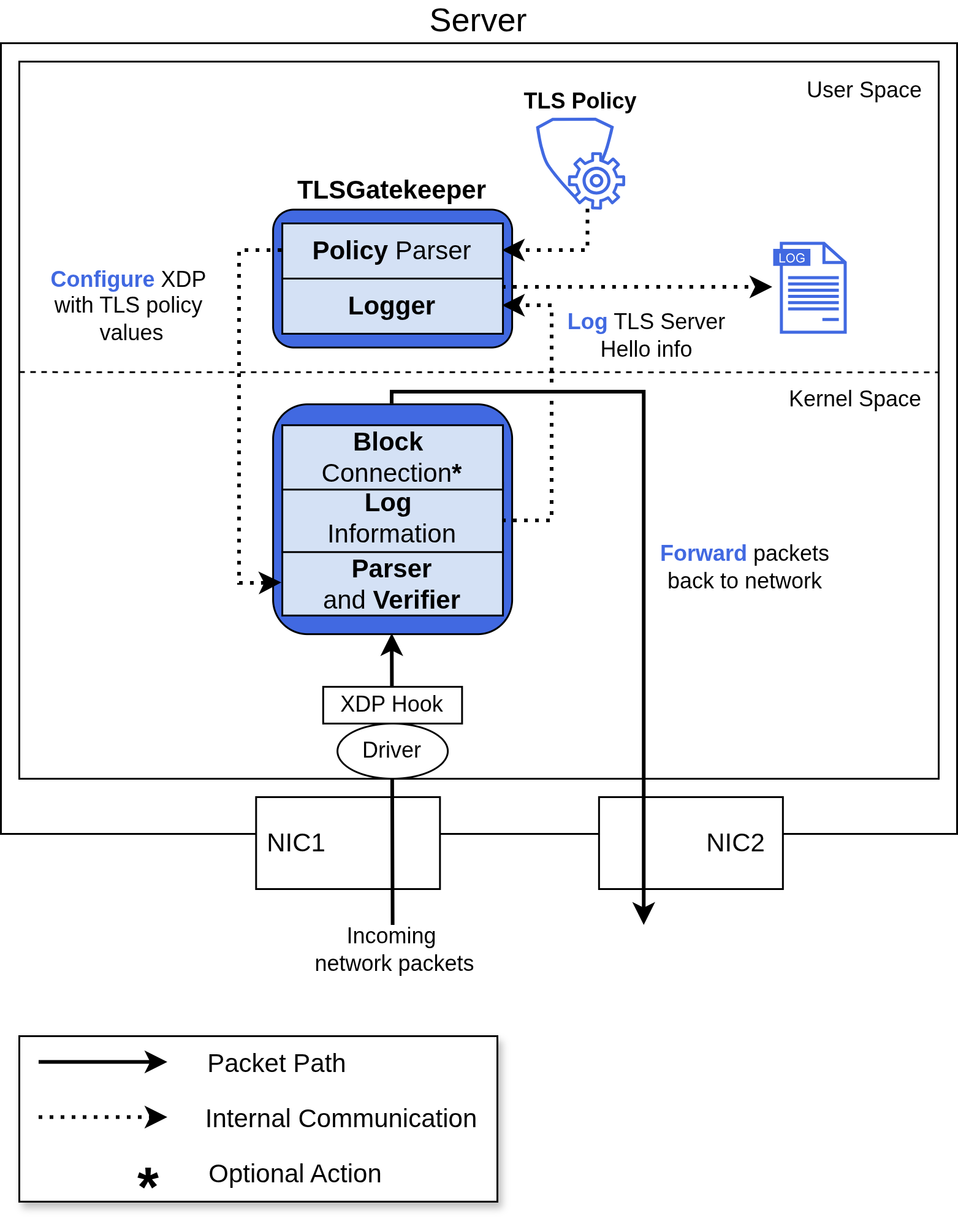}
    \caption{\toolname~elements and internal logic.}
    \label{fig:tlsgatekeeper}
\end{figure}

Middleboxes must process packets in real time and at the line rate of the monitored link. To achieve this goal, we used XDP \cite{ebpf2025,hoiland2018express} to implement \toolname. By doing so, \toolname~can process packets as soon as the driver receives them without undergoing the delay of the network stack, allowing for high-speed packet processing. The internal logic of \toolname~is illustrated in Fig. \ref{fig:tlsgatekeeper}. There are two main components in \toolname: a user space program and an XDP (kernel space) program. The user space program receives a \ac{tls} policy file, generated by a conversion script detailed in Appendix \ref{apx:conversion}, which specifies the allowed \ac{tls} versions,~\ciphers, and supported groups by the policy. 

Based on this policy file, the user space program updates two arrays in the XDP code that indicate the compliant \ac{tls} versions and \ac{tls}~1.3~\ciphers. Static arrays were chosen for these parameters because they contain, at most, two and five values, respectively. For the compliant \ac{tls}~1.2~\ciphers~and supported groups, which may include a larger number of entries, two \texttt{array} maps are used as bitmaps of $2^{16}$ positions, each one representing all possible~\ciphers~and supported groups. Compliant entries in these bitmaps are set to one. The choice of using \texttt{array} maps to create bitmaps, instead of using the more direct \texttt{hash} map, was based on the performance results presented by~\cite{liu2024understanding}. Finally, the logger uses a \texttt{ringbuf} map for efficient data streaming between the kernel and user space, waits for any log messages, and saves them into a log file. 

%  The authors reported that array map lookups are significantly faster than hash map lookups across different configurations and cache scenarios. Since this work prioritizes speed over memory, and the memory cost of both array maps is just 131 KB (two maps with $2^{16}$ positions of one byte each), this trade-off is acceptable.

After this initial configuration stage, the XDP program is attached to the target network interface and performs three operations for every incoming packet: (1) it parses the packet, and if it contains a TLS server handshake message (i.e., \emph{Server Hello} or \emph{SKE}), retrieves the relevant policy parameters and verifies compliance (all other packets simply follow their network path at this stage); (2) it logs information about the server’s selected parameters; and (3) optionally, it blocks the TLS connection if it is non-compliant. Finally, packets continue through the network, allowing them to reach their intended destination.

\subsubsection{Parser and Verifier}\label{sec:parser}

The initial stage involves parsing incoming packets, extracting the TLS server-selected parameters, and verifying their compliance with the chosen policy. In our implementation, \toolname~supports only traditional TLS-over-TCP connections and does not support DTLS or QUIC. Once the TCP header is parsed and the payload is reached, the parser follows one of two paths. The first path is for new connections and involves parsing the \emph{Server Hello} message. The second path applies only to ongoing \ac{tls}~1.2 connections where the \emph{Server Hello} was already processed, but the \emph{SKE} message (and hence the supported group) was not found in that packet.

The default parsing executed for every new connection begins when the packet containing the \emph{Server Hello} message is located. Once it's found, the parser retrieves the tentative version and the \cipher from it. If the tentative version is not \ac{tls}~1.2 (\ac{tls}~1.3 also uses version 1.2 for backward compatibility), the connection is automatically deemed non-compliant due to the use of a deprecated or non-standardized version, and the next task, logging, starts. Otherwise, parsing continues and the \emph{Server Hello} extensions are analyzed. If the \emph{Server Hello} corresponds to \ac{tls}~1.3, the parser identifies the actual version in the \texttt{supported\_versions} extension and the supported group in the \texttt{key\_share} extension. For \ac{tls}~1.3, this completes the parsing process, as all required TLS parameters have been extracted.

However, if the \texttt{supported\_versions} extension is not present, the protocol version is confirmed to be 1.2. In this case, parsing continues, since the supported group is not included in the \emph{Server Hello} but rather in the \emph{SKE} message. \toolname’s parser then proceeds either within the same handshake record,\footnote{TLS is composed of records, each containing one or more messages. For example, a TLS record of type \emph{Handshake} may include messages such as \emph{Client Hello}, \emph{Server Hello}, or \emph{Certificate}.} inspecting subsequent messages, or across additional records until it locates the \emph{SKE} message and retrieves the supported group. Since this message typically appears after the \emph{Certificate} and occasionally after further messages, it often ends up in a subsequent packet due to TCP segmentation. To handle this, \toolname~tracks segmented server messages using a \texttt{LRU hash} map, which stores the flow's five-tuple and the correct byte offset from which to resume parsing in following packets.

As \toolname~receives subsequent packets from the monitored flows, it follows the second parsing path. The parser resumes from the position stored in the map, updating the offset with each newly processed non-\emph{SKE} message, until the \emph{SKE} message is located, either in the packet immediately after the \emph{Server Hello} or in a later packet. If the \emph{SKE} message is eventually located (some \ac{tls}~1.2 handshakes do not include it), the supported group is extracted, and \toolname~has successfully gathered all required parameters. If the \emph{SKE} message is never found or does not use the \texttt{named\_curve} option, the supported group is set to zero. 

Once all the required TLS parameters are collected, \toolname~verifies compliance by sequentially checking them against the pre-loaded static arrays and eBPF maps containing the policy-approved values. The verification process begins with the protocol version, followed by the~\ciphers, and finally the supported groups. If any of the retrieved values are not found in these data structures, the packet is classified as non-compliant and the verification finishes. Subsequently, \toolname~logs the corresponding server options and compliance result.

\subsubsection{Log Information}

Once the server options are classified as compliant or not, the XDP program sends logging data to the user space logger. It creates a log message containing the following information: VLAN ID, Layer 3 (network) protocol, Layer 4 (transport) protocol, source and destination IPs, source and destination ports, TLS version, cipher suite, supported group, and the compliance outcome of the server (including the reason, if non-compliant). The set of fields included in the message can be easily modified or expanded to support additional information. Once assembled, the message is submitted to the user space program via the \texttt{ringbuf} map. The logger polls for these messages and stores them in a log file for later analysis.

\subsubsection{Blocking TLS Connections}\label{sec:blocking}

\toolname~can prevent TLS connections that are classified as non-compliant from materializing. It does so by converting the last packet in which all server parameters were analyzed into a TLS \emph{Alert}, indicating a handshake failure. This packet can correspond to the initial \emph{Server Hello}, a subsequent one containing the \emph{SKE} message, or the final server handshake packet. As a result, instead of receiving the server’s handshake messages, the client receives the alert and immediately terminates the connection with the server. To modify this packet, \toolname~uses the eBPF helper function \texttt{bpf\_xdp\_adjust\_tail} to first remove the existing TCP payload and then append the \emph{Alert} message, which is only seven bytes long. As a final step, the IP and TCP checksums are recalculated to ensure the packet is not dropped and that the client receives the packet to terminate the connection. 

\subsection{Performance Analysis}\label{sec:performance}

\begin{figure}
    \centering
    \includegraphics[width=\linewidth]{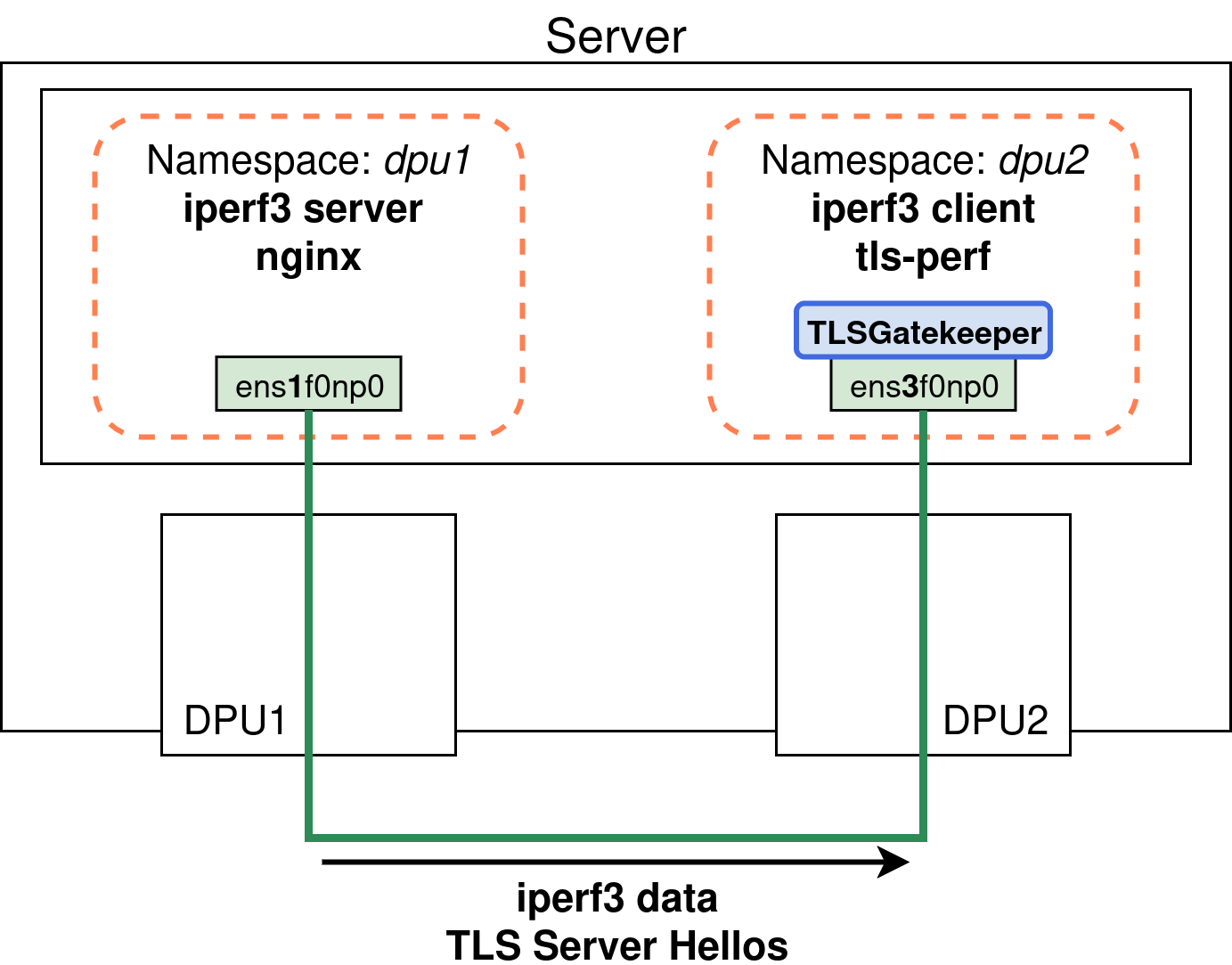}
    \caption{Performance experiments setup.}
    \label{fig:performance_setup}
\end{figure}

Since \toolname~acts as a middlebox, it must handle high-volume network traffic without becoming a bottleneck. To assess its performance,
we evaluated \toolname~in a 100 Gbps network with sustained line-rate traffic. The setup for the performance experiments is shown in Fig. \ref{fig:performance_setup}.

We connected two BlueField-2 DPUs (or SmartNICs) using a 100 Gbps link and connected them to the same server. The server contains an AMD EPYC 9224 CPU with 24 cores and 48 threads. Both DPUs operated in NIC (passthrough) mode. Since both DPUs were within the same server, we created two network namespaces (i.e., \textit{ip netns}), one for each DPU, to enable communication over the 100 Gbps link. In the experiments, DPU1 represents the external network, where the TLS servers are located, while DPU2 represents the internal network, where TLS clients initiate connections to the servers. \toolname~is positioned on the interface connected to DPU2 and processes the incoming packets from DPU1.

\begin{figure*}
     \centering
     \begin{subfigure}[b]{0.31\textwidth}
         \centering
         \includegraphics[width=\textwidth]{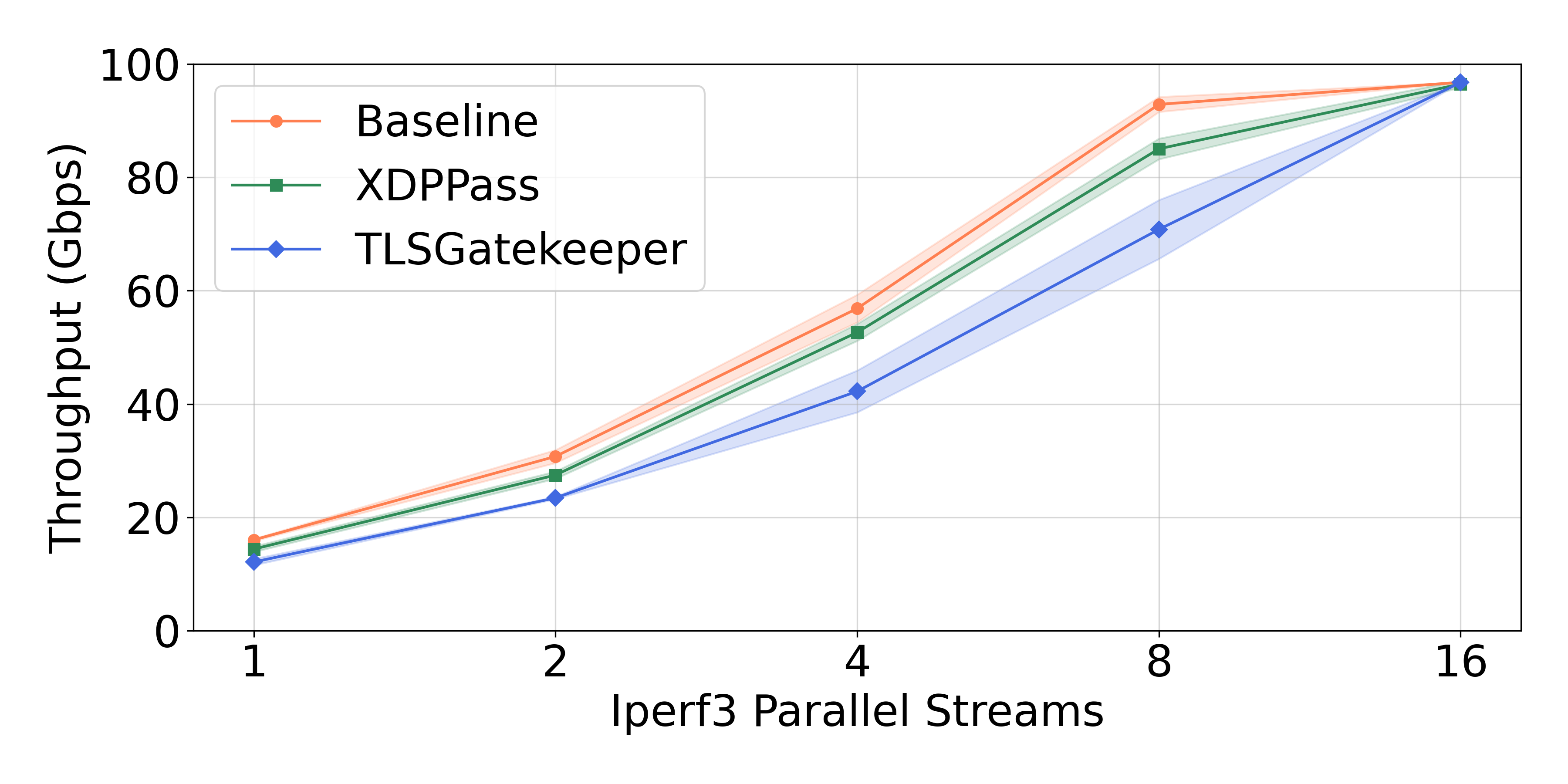}
         \caption{Throughput with increasing \textit{iperf} streams.}
         \label{fig:throughput}
     \end{subfigure}
     \hfill
     \begin{subfigure}[b]{0.31\textwidth}
         \centering
         \includegraphics[width=\textwidth]{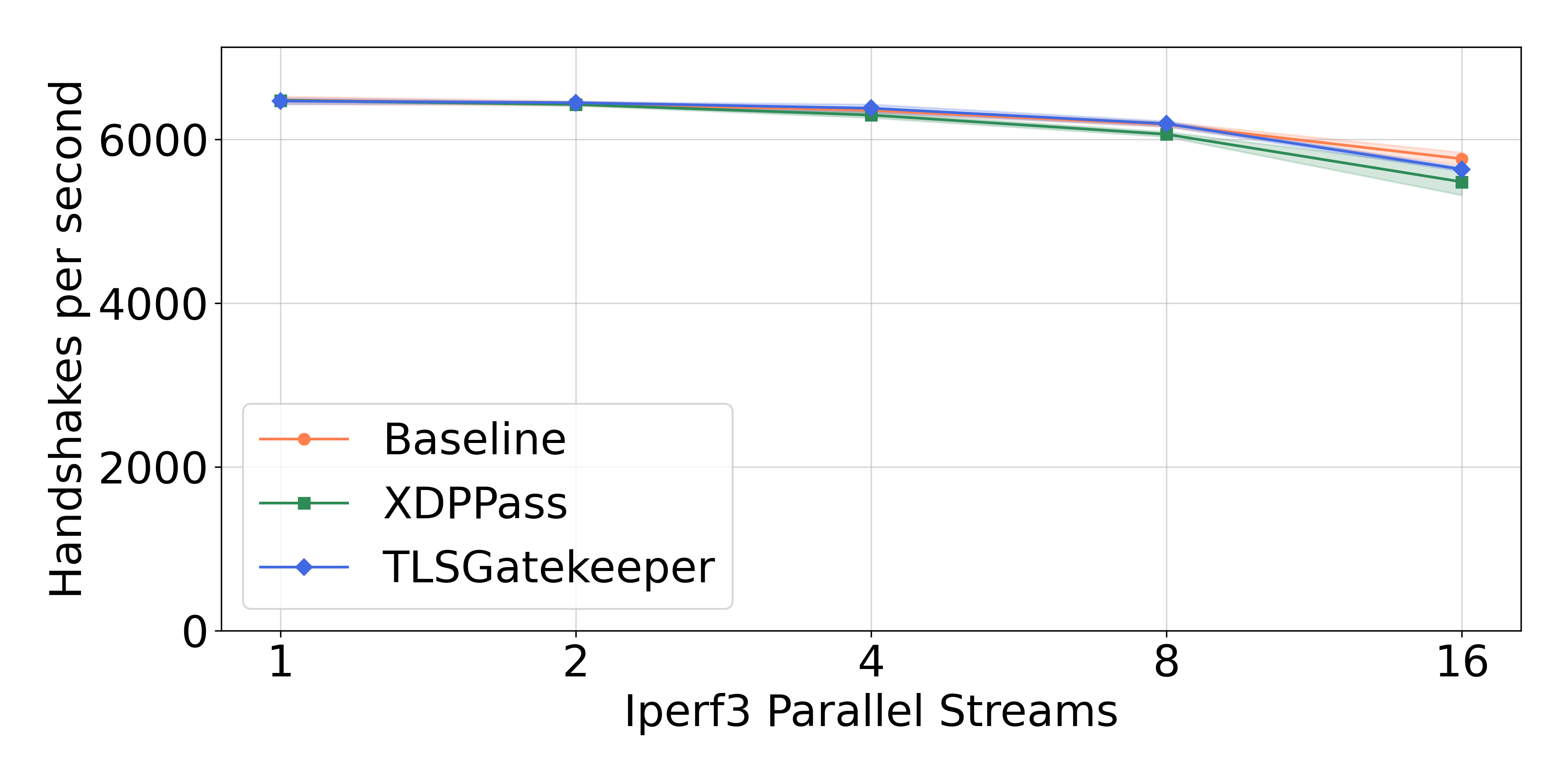}
         \caption{95th percentile of TLS handshakes per second with increasing \textit{iperf} streams.} % (95\% of measurements are \textbf{above} this value)}
         \label{fig:handshakes-sec}
     \end{subfigure}
     \hfill
     \begin{subfigure}[b]{0.31\textwidth}
         \centering
         \includegraphics[width=\textwidth]{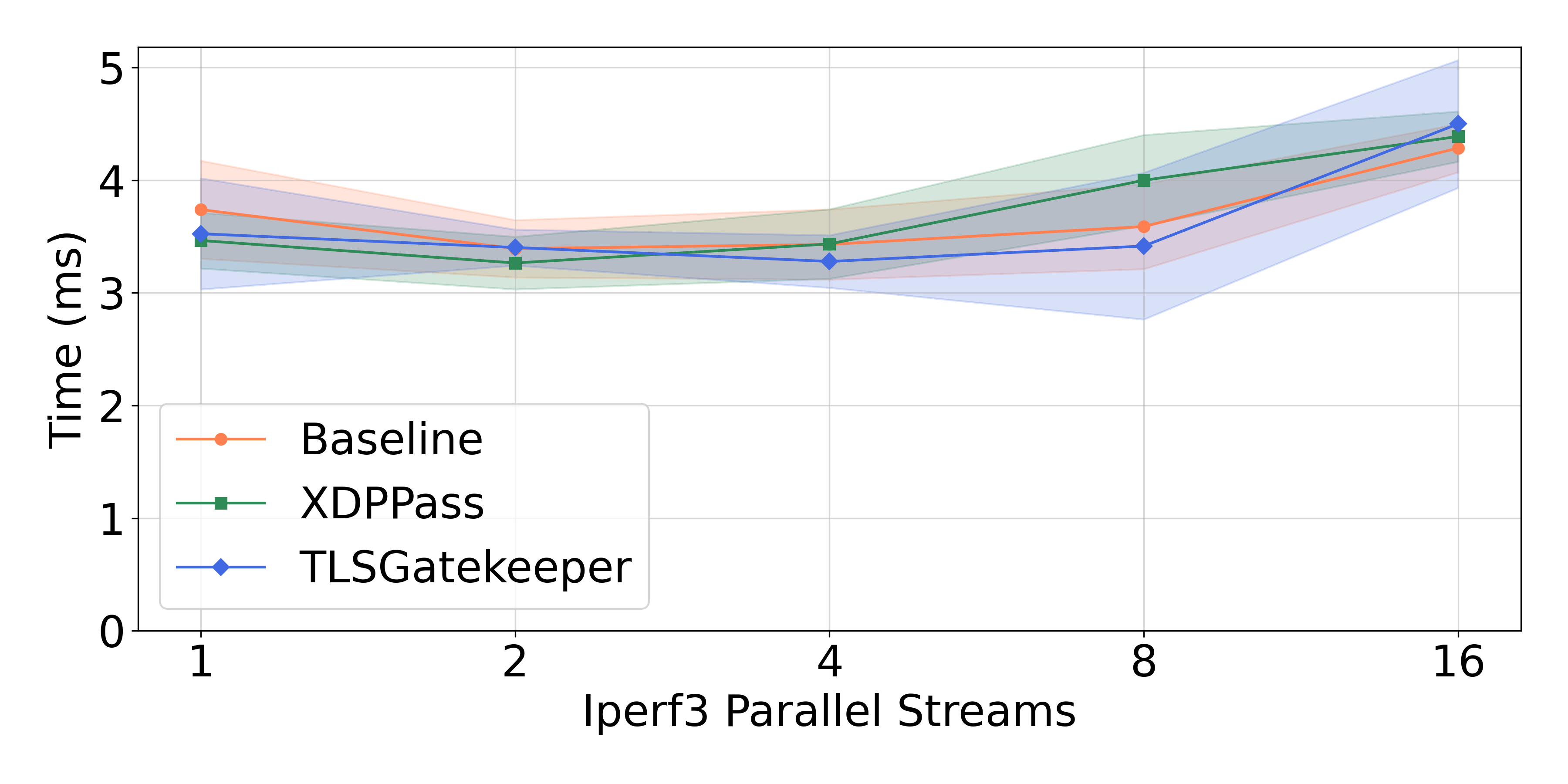}
         \caption{95th percentile of TLS latency with increasing \textit{iperf} streams.} % (95\% of measurements are \textbf{below} this value)}
         \label{fig:latency}
     \end{subfigure}
        \caption{Performance comparison of~\toolname~against two reference scenarios: a clean interface with no programs attached (Baseline) and an interface running a minimal XDP program (XDPPass).}
        \label{fig:performance-graphs}
\end{figure*}

For the experiment's network traffic, we used \textit{iperf3}\footnote{\url{https://github.com/esnet/iperf}} to generate background TCP traffic and \textit{tls-perf}\footnote{\url{https://github.com/tempesta-tech/tls-perf/tree/master}} to initiate TLS connections. In the DPU1 namespace, we set up an \textit{iperf3} server and an \textit{nginx}\footnote{\url{https://nginx.org/}} server, while in the DPU2 namespace, we ran the \textit{iperf3} client with the reverse option enabled, which means that the server generates traffic for the client, and executed \textit{tls-perf} to start TLS connections with the \textit{nginx} server. With this setup, \toolname~receives both the TCP background traffic generated by the \textit{iperf3} server and the TLS handshake messages sent by the \textit{nginx} server.

Regarding \toolname, we configured it to not block TLS traffic, allowing the measurement of handshake metrics, and created a policy based on the NIST guideline. For this setup, \toolname’s implementation passes all packets to the network stack via \texttt{XDP\_PASS}, allowing them to reach the Linux operating system where the clients are running. In the typical middlebox deployment presented in Fig. \ref{fig:deployment}, the XDP program would instead use \texttt{XDP\_REDIRECT} to forward packets to another NIC and back to the network. Additionally, during these experiments, the XDP component of \toolname~measures and logs the processing delay incurred while parsing server messages. To minimize logging overhead, this delay is recorded only for packets that had a compliance outcome (i.e., TLS \emph{Server Hellos}) and, inevitably, must send a log to user space. Although it does not represent the delay experienced by all packets in the network, these measurements are the most relevant, as such packets undergo the most extensive processing by our tool.

Finally, we compared \toolname~against two other scenarios: a \textit{Baseline} with no XDP code running at all, and another called \textit{XDPPass} where a simple XDP program merely passes packets to the kernel.

\subsubsection{Reaching 100 Gbps Throughput}

To evaluate how \toolname~handles increasing traffic volume, we conducted experiments using \textit{iperf3} to approach the 100 Gbps bandwidth limit. Specifically, we varied the number of parallel \textit{iperf3} streams from 1 to 16, following a power-of-two sequence. For TLS traffic, \textit{tls-perf} was configured to use 10 threads with two parallel connections per-thread to generate a high volume of \ac{tls} connections. The resulting number of handshakes per second in these tests significantly exceeds the maximum rate observed in our institution's traffic (see Fig. \ref{fig:handshakes-sec}), which was roughly 200 handshakes per second. Finally, for each \textit{iperf3} stream amount and scenario, we conducted five runs to reduce the impact of outliers.

Fig.~\ref{fig:throughput} shows the average RX throughput on the DPU2 interface across five runs as the number of \textit{iperf3} streams increases up to 16. With 16 streams, \textit{iperf3} reaches nearly 100~Gbps, which is the maximum throughput supported by the evaluated DPUs. Although \toolname~shows a lower throughput than the \textit{Baseline} and \textit{XDPPass} with fewer parallel streams, it maintains full performance once the link becomes saturated with 16 streams, demonstrating that \toolname~can handle high-volume throughput.

Regarding TLS-related metrics, Fig. \ref{fig:handshakes-sec} shows the 95th percentile of handshakes per second, while Fig. \ref{fig:latency} presents the 95th percentile of handshake latency as reported by \textit{tls-perf}. For handshakes per second, the 95th percentile indicates that 95\% of the measurements were above that value; conversely, for latency, it means that 95\% of the TLS handshake latency measurements were below the indicated value. The plots represent the average 95th percentile over the five runs. Across all scenarios, we observe the expected trend of decreasing handshakes per second and increasing latency as background traffic rises. Both graphs show that \toolname~does not cause any noticeable difference compared to the other scenarios, demonstrating its ability to handle thousands of TLS handshakes per second with up to 100~Gbps of background traffic, without disrupting the expected network behavior.

\begin{figure}
    \centering
    \includegraphics[width=\linewidth]{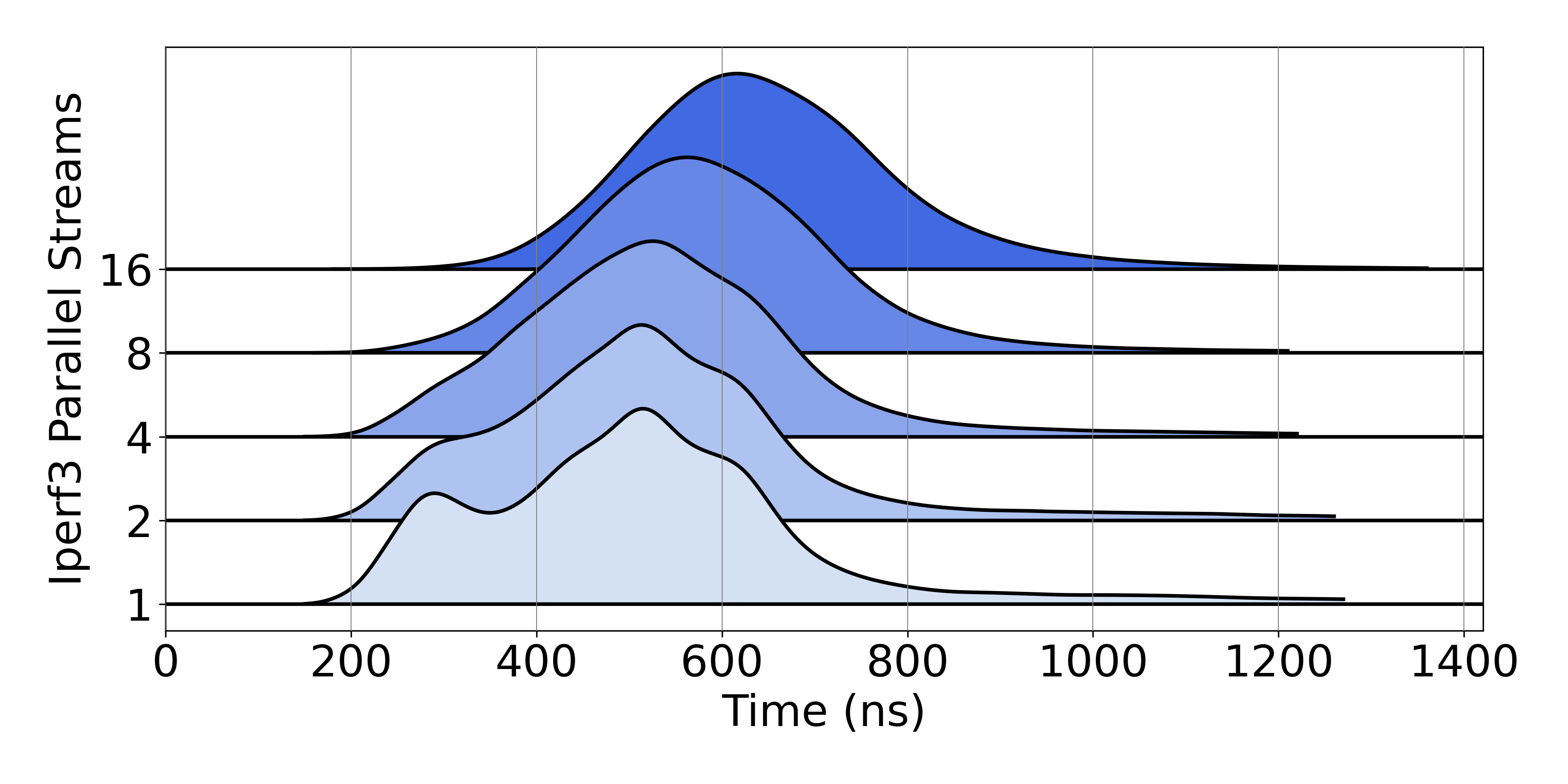}
    \caption{\toolname~processing delay up to the 99th percentile.}
    \label{fig:tlsgatekeeper-delay}
\end{figure}

\toolname's minimal impact on network performance is due to its contained processing, as quantified by Fig.~\ref{fig:tlsgatekeeper-delay}. It shows the packet processing delay introduced by~\toolname, from the packet's arrival to the compliance outcome. This delay is only measured for packets that trigger a compliance check (i.e. packets with a \emph{Server Hello} message or \emph{SKE}) and represents the maximum overhead added to any single packet. Ridgeline plots in the figure present the delay distribution across all five runs for each \textit{iperf3} stream configuration. The results indicate that, even if the processing delay slightly increases under higher traffic loads (more parallel streams), the overhead remains negligible compared to the TLS latency shown in Fig.~\ref{fig:latency} (milliseconds versus nanoseconds).

\subsubsection{\ac{tls}~1.2 vs \ac{tls}~1.3}

The previous results focused on \ac{tls}~1.3, which has a simpler structure and is therefore easier for \toolname~to parse. To evaluate the effect of the additional parsing logic required for \ac{tls}~1.2 (described in Section \ref{sec:parser}), we repeated the experiments using \textit{tls-perf} configured for that version. We then compared both protocols under a load of 16~\textit{iperf3} streams, corresponding to nearly 100~Gbps of background traffic.

\begin{figure}
    \centering
    \includegraphics[width=\linewidth]{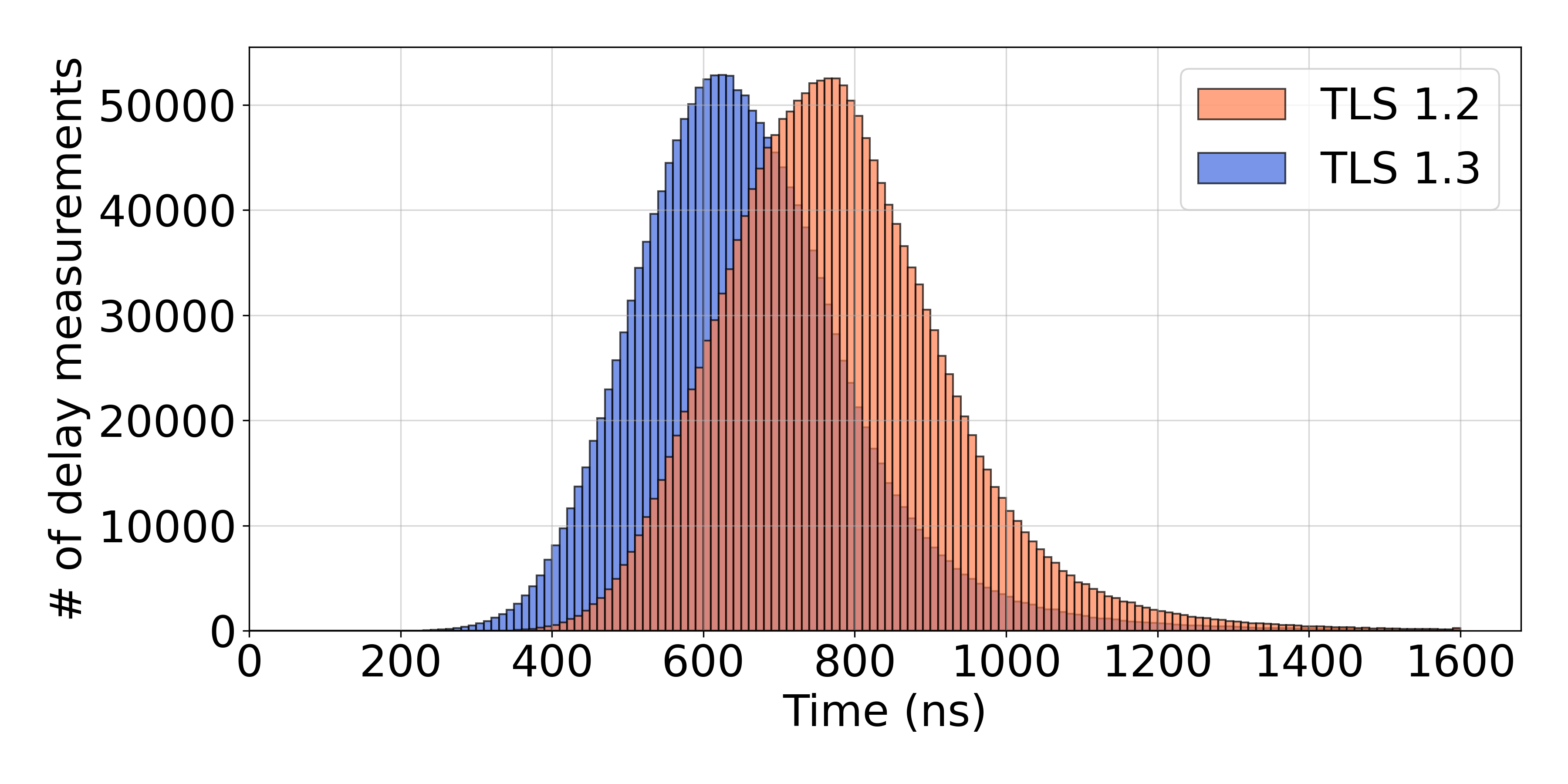}
    \caption{Comparing \toolname~delay for \ac{tls}~1.2 and 1.3 up to the 99th percentile.}
    \label{fig:tlsgatekeeper-delay-hist}
\end{figure}

The throughput remained virtually identical in both versions, with every scenario reaching close to 100~Gbps. The number of handshakes per second was slightly lower for \ac{tls}~1.3 in all scenarios, which was unexpected but has been previously observed in OpenSSL.\footnote{\url{https://github.com/openssl/openssl/issues/13053}} Latency remained similar between the two versions, with a minor increase for \toolname~when processing \ac{tls}~1.2 traffic. This increase is attributed to the additional parsing overhead introduced by the tool. Fig.~\ref{fig:tlsgatekeeper-delay-hist}~quantifies this effect, showing the distribution of processing delays across five runs with 16~\textit{iperf3} streams. The average delay from all these measurements was 671 ns for \ac{tls}~1.3 and 795 for \ac{tls} 1.2. Despite this added cost, the delay remains negligible relative to the overall handshake latency (see Fig.~\ref{fig:latency}), confirming that \toolname~operates efficiently for both protocol versions.

\subsection{Blocking PoC}\label{sec:blocking_poc}

\toolname~can block non-compliant TLS connections as detailed in Section \ref{sec:blocking}. In the performance experiments of Section \ref{sec:performance}, this feature was disabled, and \toolname~only logged TLS-related information, as the goal was to evaluate handshake throughput and latency. To demonstrate \toolname’s gatekeeping capability, this section presents a small Proof-of-Concept (PoC) experiment involving two scenarios. In the first scenario, neither the server nor the client complies with the desired policy on supported groups, causing \toolname~to prevent the connection from materializing. In the second scenario, the server is compliant with the policy but the client is initially not (it's default supported group is not approved by the policy); however, since it is the server that ultimately determines the cryptographic options, the clients updates it's supported groups, and \toolname~allows the connection to proceed.

The setup is the same as in Fig.~\ref{fig:deployment} but without the \textit{iperf3} tool. For \toolname, the NIST-based policy is applied with the blocking option enabled. In the first scenario, the \textit{nginx} server is configured without restricting supported groups, thereby accepting all groups allowed by OpenSSL. The \textit{tls-perf} client initiates five TLS 1.3 connections proposing the supported group \texttt{Curve25519}, which the TLS server accepts. Since \texttt{Curve25519} is not permitted by the NIST-based policy, \toolname~blocks the connections. Once the experiment starts, \toolname~intercepts the \emph{Server Hello} containing the non-compliant supported group, modifies it into a TLS \emph{Alert}, and forwards it to the client. As shown in Listing \ref{lst:blocking}, when the client receives the alert indicating a handshake failure, it terminates the connection, causing the client application to close. This entire process occurs transparently within the network and requires no client-side configuration.

\begin{figure}[htpb]
\begin{lstlisting}[escapechar=!]
Running TLS benchmark with following settings:
Host:        192.168.10.10 : 7777
TLS version: 1.3
Cipher:      default
TLS tickets: off
Duration:    0

( All peers are active, start to gather statistics )
!\textcolor{red}{ERROR: cannot establish even one TLS connection: error:0A000410:SSL routines::sslv3 alert handshake failure}
\end{lstlisting}
\captionof{lstlisting}{\textit{tls-perf} log when TLSGatekeeper is \textbf{blocking} the connection.}
\label{lst:blocking}
\end{figure}

\begin{figure}[htpb]
\begin{lstlisting}
Running TLS benchmark with following settings:
Host:        192.168.10.10 : 7777
TLS version: 1.3
Cipher:      default
TLS tickets: off
Duration:    0

( All peers are active, start to gather statistics )
TLS hs in progress 0 [5 h/s], TCP open conns 0 [1 hs in progress], Errors 0
========================================
 TOTAL:           SECONDS 1; HANDSHAKES 5
 HANDSHAKES/sec:  MAX 5; AVG 5; 95P 5; MIN 5
 LATENCY (ms):    MIN 1.73175; AVG 1.82172; 95P 2.28168; MAX 2.28168
\end{lstlisting}
\captionof{lstlisting}{\textit{tls-perf} log when TLSGatekeeper is \textbf{not blocking} the connection.}
\label{lst:not_blocking}
\end{figure}

In the second scenario, the \textit{nginx} server is configured to use the supported group \texttt{prime256v1}, which is approved by NIST and represents a compliant TLS server. As before, the \textit{tls-perf} client proposes \texttt{Curve25519}, which NIST does not accept. Since the server ultimately determines the final values and, in this scenario, is configured to use \texttt{prime256v1}, the \emph{Server Hello} message contains a compliant value. Consequently, \toolname~does not block the connection. As shown in Listing \ref{lst:not_blocking}, \textit{tls-perf} successfully completes the handshakes, and \toolname~simply logs the session as compliant.

While blocking may appear intrusive, it provides strong protection by preventing connections to potentially insecure servers, and allows system administrators to review contacted servers without haste.
\section{Conclusion}\label{sec:conclusions}

In this paper, we analyzed over 50 million TLS handshakes collected over a two-week period to evaluate how real-world server parameters align with the latest security guidelines. Our analysis yielded two main findings. First, although the use of outdated versions and insecure \ciphers~is minimal, their continued usage is problematic. Second, servers are adopting the latest TLS developments faster than security guidelines are updated to address them. The dataset and our analysis are publicly available to the research community.

To protect clients from insecure servers and provide organizations with a reliable tool to enforce TLS policies, we introduced \toolname. By monitoring only TLS handshakes and blocking connections that use non-compliant options, \toolname~operates transparently without requiring client-side modifications. Our performance evaluation demonstrates that \toolname~effectively monitors and filters connections under a 100 Gbps traffic load while introducing negligible latency.
{\appendices

\section{TLS Compliance Dataset}\label{apx:tls_guidelines}

The TLS Compliance Dataset is a publicly auditable dataset containing technical requirements extracted from multiple guidelines written by government agencies. The dataset is composed of twelve Markdown files (one for each set of configurable elements) containing a table that links each element with an RFC2119-defined requirement level. Additionally, some of the guidelines define supplemental conditions that restrict the usage of the configurations; these cases are reported in a dedicated column. Finally, machine-readable formats of the dataset are provided by the authors, namely in OpenDocument Spreadsheet (ods) and SQLite. 

\section{Guideline to Policy Conversion Script}\label{apx:conversion}

As shown in Figure~\ref{fig:tlsgatekeeper},~\toolname~receives a ``TLS Policy'' file specifying the policy's accepted protocol versions,~\ciphers, and supported groups. To facilitate using recommended values from existing TLS guidelines, we created a fork from \tlsa~and developed a new module for it.\footnote{Stored in the same repository as the dataset.} This module builds upon the existing compliance component to create a binary file containing all elements allowed by a given guideline from the aforementioned parameters. The allowed elements for each guideline are derived from the~\datasetname~by selecting those with a requirement level of \textit{MUST}, \textit{RECOMMENDED}, or \textit{OPTIONAL}\footnote{The dataset’s authors use the term “requirement level” to denote the assignment of a keyword from RFC2119 to an element.}.

In this binary file, the policy's valid elements are divided into four categories: (1) protocol versions, (2)~\ciphers~supported by \ac{tls}~versions up to 1.2, (3)~\ciphers~supported by \ac{tls}~1.3, and (4) supported groups. Since, at the time of writing, \ac{tls}~1.3 supports only seven~\ciphers~compared to over one hundred available in version 1.2, the cipher categories were separated to improve performance when handling \ac{tls}~1.3 packets. For each category, one byte is used to identify the category, two bytes specify the length of the data within that category, and each element is mapped to its corresponding identifier using the values from the IANA registry~\cite{tlsparameters}. This file provides a simple input for~\toolname~to identify non-compliant connections.

\bibliographystyle{IEEEtran}
\bibliography{bibliography}

% \newpage

% \section*{Biography Section}
 

% \end{IEEEbiography}



% \vfill

\end{document}